\documentclass[fleqn,usenatbib]{mnras}

\usepackage{newtxtext,newtxmath}
\usepackage{comment}
\usepackage[T1]{fontenc}
\usepackage[utf8]{inputenc}
\DeclareRobustCommand{\VAN}[3]{#2}
\let\VANthebibliography\thebibliography
\def\thebibliography{\DeclareRobustCommand{\VAN}[3]{##3}\VANthebibliography}

\usepackage{graphicx}	
\usepackage{amsmath}	
\usepackage{physics}
\usepackage{url}
\usepackage{siunitx}

\defcitealias{2015MNRAS.446.3297K}{K15}

\usepackage{xcolor}
\definecolor{cadmiumgreen}{rgb}{0.0, 0.42, 0.24}
\definecolor{softblue}{RGB}{0, 50, 175}
\newcommand{\pk}[1]{\textcolor{softblue}{#1}} 
\newcommand{\cat}{\ion{Ca}{II}}

\title[Pal 5 Tidal Tails]{Forward and Back: Kinematics of the Palomar 5 Tidal Tails}

\author[P. B. Kuzma et al.]{P. B. Kuzma,$^{1}$ \thanks{E-mail: pkuzma@roe.ac.uk (PK)}
A. M. N. Ferguson$^{1}$,
A. L. Varri$^{1,2}$,
M. J. Irwin$^{3}$,
E. J. Bernard$^{4}$,
\newauthor
E. Tolstoy$^{5}$,
J. Pe\~{n}arrubia$^{1}$,
D. B. Zucker$^{6,7}$
\\
$^{1}$Institute for Astronomy, University of Edinburgh, Royal Observatory, Blackford Hill, Edinburgh, EH9 3HJ, UK\\
$^{2}$School of Mathematics and Maxwell Institute for Mathematical Sciences, University of Edinburgh, King's Buildings, Edinburgh EH9 3FD, UK\\
$^{3}$ Institute of Astronomy, University of Cambridge, CB3 0HA, UK\\
$^{4}$ Universit\'e C\^ote d'Azur, OCA, CNRS,  Lagrange, Boulevard de l'Observatoire, CS 34229, F-06304 Nice cedex 4, France\\
$^{5}$ Kapteyn Astronomical Institute, University of Groningen, Postbus 800, 9700AV Groningen, The Netherlands\\
$^{6}$ Department of Physics and Astronomy, Macquarie University, Balaclava Road, Sydney, NSW 2109, Australia\\
$^{7}$ Research Centre for Astronomy, Astrophysics and Astrophotonics, Macquarie University, Balaclava Road, Sydney, NSW 2109 Australia}

\date{Accepted XXX. Received YYY; in original form ZZZ}

\pubyear{2021}

\begin{document}
\label{firstpage}
\pagerange{\pageref{firstpage}--\pageref{lastpage}}
\maketitle

\begin{abstract}
The tidal tails of Palomar 5 (Pal 5) have been the focus of many spectroscopic studies in an attempt to identify individual stars lying along the stream and characterise their kinematics. The well-studied trailing tail has been explored out to a distance of 15$\degr$ from the cluster centre, while less than four degrees have been examined along the leading tail. In this paper, we present results of a spectroscopic study of two fields along the leading tail that we have observed with the AAOmega spectrograph on the Anglo-Australian telescope. One of these fields lies roughly 7$\degr$ along the leading tail, beyond what has been previously been explored   spectroscopically. Combining our measurements of kinematics and line strengths with Pan-STARRS1 photometric data and Gaia EDR3 astrometry, we adopt a probabilistic approach to identify 16 stars with high probability of belonging to the Pal 5 stream. Eight of these stars lie in the outermost field and their sky positions confirm the presence of ``fanning'' in the leading arm. We also revisit previously-published radial velocity studies and incorporate Gaia EDR3 astrometry to remove interloping field stars.   With a final sample of 109 {\it bona fide} Pal 5 cluster and tidal stream stars, we characterise the 3D kinematics along the the full extent of the system. We provide this catalogue for future modeling work.
\end{abstract}

\begin{keywords}
stars: abundances -- stars: kinematics and dynamics: -- globular clusters: general -- globular clusters: individual: Palomar 5 
\end{keywords}


\section{Introduction}
The number of Milky Way (MW) globular clusters (GCs) known to possess extended tidal features is growing. Thanks to large, well-calibrated imaging surveys such as Sloan Digital Sky Survey (SDSS) \citep{2012ApJS..203...21A} and Pan-STARRS1 (PS1) \citep{2016arXiv161205560C}, as well as the more recent astrometric Gaia Space Mission \citep{2018A&A...616A...1G}, tidal features have been found around a few tens of GCs in the MW halo \citep[e.g.,][Kuzma et al., in prep]{2020MNRAS.495.2222S}, ranging from the massive inner halo NGC 5139 \citep{2019NatAs.tmp..258I,2021MNRAS.507.1127K} to the low mass outer halo Palomar 1 \citep{2010MNRAS.408L..66N}. The most emphatic display of tidal features in the MW, however, belongs to Palomar 5 \citep[Pal 5,][]{2001ApJ...548L.165O,2003AJ....126.2385O}. The tidal tails of Pal 5 have now been traced across more than 20$\degr$ on the sky \citep{2006ApJ...641L..37G, 2016MNRAS.463.1759B, 2020ApJ...889...70B} and it remains one of the very few stellar streams in the halo with a known progenitor. 

As the most striking example of GC disruption in action, the tidal tails of Pal 5 have been the subject of many studies over the last two decades, many of which have focused on how their properties can constrain the Galactic potential  \citep[e.g.,][]{2004AJ....127.2753D,2012A&A...546L...7M,2014ApJ...795...94B,2015ApJ...803...80K,2015ApJ...799...28P, 2016ApJ...833...31B}. The present-day stellar mass of Pal 5 plus its tidal tails has been estimated to be $1.2-9\times 10^4$ M$_{\odot}$, with 4300 M$_{\odot}$ of that retained in the main body \citep{2017ApJ...842..120I, 2019AJ....158..223P}. This implies that roughly half of the total mass of the system is now populating the tails. The tails also show some level of substructure, including gaps and peaks \citep[e.g.][]{2012ApJ...760...75C, 2016ApJ...819....1I, 2017MNRAS.470...60E}, ``wiggles" and ``fanning" \citep{2020ApJ...889...70B}.  

While the reality and origin of fine structure along the Pal 5 tails has often been debated \citep[e.g.,][]{2016MNRAS.460.2711T, 2016ApJ...819....1I}, most studies agree on the fact that there is a gross asymmetry in the length of the leading and trailing tails.  Early mapping work was limited by the SDSS footprint but \cite{2016MNRAS.463.1759B} were able to use the more extensive coverage of the PS1 survey to show that the leading tail could only be traced photometrically for about half the length ($\approx 8\degr$) of the trailing tail ($\approx 16\degr$). In particular, they showed that, at the photometric depth of PS1,  the leading tail could be followed to $\delta\approx-6\degr$  on the sky before ending abruptly.  Using deeper photometry from the DECam Legacy Survey \cite[DECaLS; ][]{2019AJ....157..168D}, \citet{2020ApJ...889...70B} demonstrated that leading tail debris could be detected slightly beyond this point, but that it was spread out in a low-density fan. Such fanning could indicate triaxiality in the potential \citep{2015ApJ...799...28P} or the effect of a rotating bar 
\citep[see][]{,2017NatAs...1..633P,2020ApJ...889...70B}.  On the other hand, 
\cite{2020MNRAS.493.4978S} analyse Gaia DR2 data and suggest that the leading tail can be traced nearly as far as the trailing tail in that dataset. However this analysis is based on main sequence turn-off stars with magnitudes at the faint limit of the survey and the assumption of a cluster metallicity of [Fe/H]=-3.1 dex compared to the measured value of [Fe/H]=-1.5 dex  \citep{2017A&A...601A..41K}.  A further complication is that, at these faint magnitudes,  there is also contamination along the sightline to the leading tail from the Sagittarius (Sgr) stream  \citep{2016ApJ...819....1I, 2020ApJ...889...70B}. 
  
Spectroscopic studies of the Pal 5 stream have thus far been more limited than photometric ones. The first kinematic exploration of the Pal 5 tails was performed by \cite{2009AJ....137.3378O}, where they explored an 8.5$^{\circ}$ extent of the tidal tails. Using 17 confirmed red giant branch (RGB) members, they measured a linear radial velocity gradient of 1.0 km s$^{-1}$ deg$^{-1}$ along the stream, as well as a small intrinsic velocity dispersion of $\sim2$ km s$^{-1}$.  \citet[][hereafter K15]{2015MNRAS.446.3297K}, \citet{2016ApJ...823..157I} and \citet[][]{2017ApJ...842..120I} (hereafter Ib17) conducted more extensive spectroscopic searches, covering a $\approx 20^{\circ}$ region along the stream, and confirmed this mild radial velocity gradient and low dispersion. 

\begin{figure*}\label{fig1}
  \begin{center}  
    \includegraphics[width=0.9\textwidth]{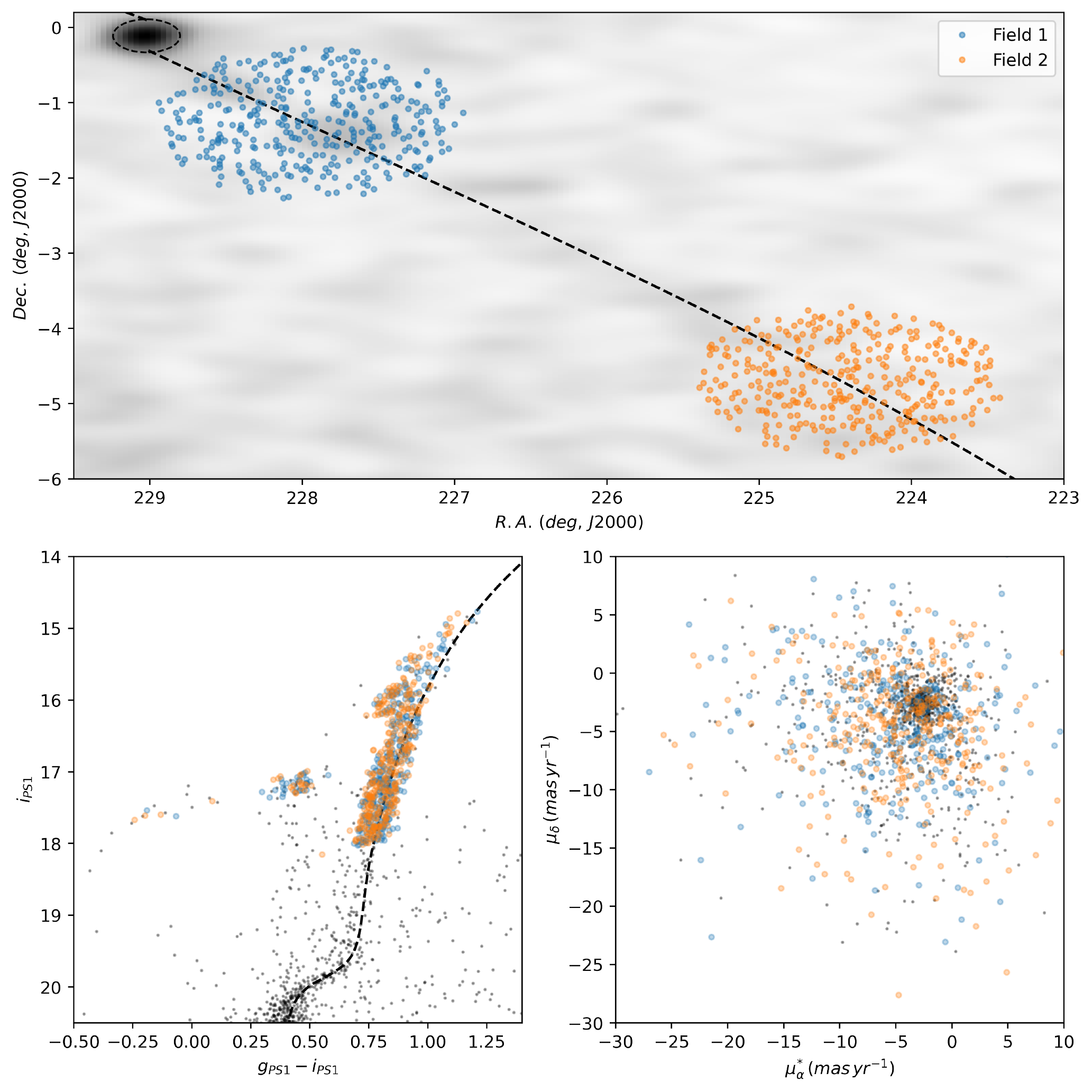}\label{fig:obs_fields}
  \end{center}
\caption{Top: Location of the observed targets overplotted on a density map of EDR3 stars. The dashed line indicates the stream track determined by \citet{2017ApJ...842..120I}. Pal 5 resides at (229$\degr$, 0$\degr$) and the dashed circle indicates the Jacobi radius of 11 arcmin \citep{2019AJ....158..223P}. Bottom: PS1 photometry (left) and Gaia EDR3 proper motions (right) of our stream targets overlaid on those of stars lying within the Jacobi radius of Pal 5. The left panel includes a Dartmouth stellar isochrone \citep{2008ApJS..178...89D} of age=13.65 Gyr and [Fe/H]=--1.56 dex. In the right panel, the  proper motion of the main body of Pal 5 is $(\mu^{*}_{\alpha},\mu_{\delta})= (-2.76,-2.65)~\rm{mas}~ \rm{yr}^{-1}$.}
\end{figure*}

While none of these radial velocity studies have found evidence for fanning in the tails, they have not probed the extreme ends of the tails where such behaviour might be expected.  \cite{2019AJ....158..223P} have recently explored kinematics along the Pal 5 streams using RR Lyrae (RRL) stars that have Gaia DR2 proper motions (PMs). Intriguingly, they find a few RRLs with high probability of stream membership to be considerably offset from the stream track, including two stars in the region where \citet{2020ApJ...889...70B} find evidence for fanning in their star count map. No radial velocities are available for these stars, however.

\begin{table}
\centering

\caption{List of observations.}
\label{tab:obslist}
\begin{tabular}{@{}ccccccc@{}}

\hline \hline
& Field 1& Field 2\\
\hline
R.A. (J2000)&$227\overset{\circ}{.}946$&$224\overset{\circ}{.}425$\\
Dec. (J2000)&$-1\overset{\circ}{.}266$&$-4\overset{\circ}{.}690$\\
Date-obs&25/02/2017&24/02/2017\\
Exp. Time&$3\times1200$s&$3\times1800$s\\
Avg. Seeing&$2.6''$&$1.5''$\\
Ang. Distance$^{\dagger}$&$1.75^{\circ}$&$6.6^{\circ}$\\
\hline
$^{\dagger}$ Angular distance from Pal 5.
\end{tabular}
\end{table}

In this paper, we present a new kinematic study of the leading Pal 5 tidal tail.  This work probes greater  angular distances from the cluster than previous radial velocity studies and includes the region where the stream has been suggested to fan out based on deep photometry. In Section 2, we discuss the observations performed and the probabilistic methods we have employed to identify stream members. Section 3  presents our results and we discuss our findings in Section 4. We present our conclusions in Section 5.

\section{The Data}
\subsection{Observations \& Target Selection}
AAOmega is a multi-fibre, dual-beam spectrograph that is mounted on the 3.9m Anglo-Australian Telescope (AAT) at Siding Springs Observatory \citep{2006SPIE.6269E..0GS}. When coupled to the Two Degree Field  (2dF) fibre positioning system \citep{2002MNRAS.333..279L},
it provides 392 science fibres that can be configured across a 2$\degr$ diameter circular field. We used AAOmega+2dF with a  dichroic centered at 5700 \r{A} to split the incoming light down the red and blue arms, which held the 1700D and 580V gratings respectively. The 1700D grating covers the wavelength range $\sim$8400--8880 \r{A}  which includes the \cat\ triplet absorption lines at 8498 \r{A}, 8542 \r{A} and 8662 \r{A} and has a resolution R=10000. In the blue, the 580V grating 
covers the range $\sim$3800--5800 \r{A}  with a resolution of R=1300 and covers the \ion{Mg}{I} triplet lines (also known as Mg $b$) at 5167 \r{A},  5173 \r{A} and 5184 \r{A}. As part of Opticon proposal 17A/063 (PI Ferguson) two fields on the Pal 5 stream were observed on 24th and 25th of February 2017 under dark conditions and mostly clear skies, with 354 and 352 stars targeted per field on respective nights.  Arc spectra and quartz lamp flat-fields were obtained before and after each set of science observations and a series of bias exposures were taken at the start of each night. 

Each target field had three sets of observations with the closest field to the main body of Pal 5 having individual exposures of 1200s while the outermost field had 1800s exposures.  Both fields were chosen to lie  along the leading tail of the Pal 5 stream as traced by \cite{2016MNRAS.463.1759B}.  Field~1 is located 1.75$^{\circ}$ from the center of the cluster, while Field~2 is at the furthest extent mapped to date, 6.6$^{\circ}$ away,  in a region that has not previously been studied spectroscopically. Fig. \ref{fig:obs_fields} shows the sky positions of our fields relative the main body of Pal 5 and the observations are summarised in Table \ref{tab:obslist}. 

The  target selection was based on the PS1 DR1 photometry  \citep{2016arXiv161205560C} with the primary consideration being the location of stars in  colour-magnitude space with respect to the expected locus for Pal~5.
In Fig. \ref{fig:obs_fields}, we show a colour-magnitude diagram (CMD) of the stars which lie within the Jacobi radius of Pal~5, calculated to be 11 arcmin by \citet{2019AJ....158..223P}, on which our spectroscopic targets are overlaid. As can be seen, our targets are fairly bright stars ({\it i$_{\rm PS1} \leq 18$}) which largely lie on the upper RGB and blue horizontal branch (HB).  
 
 As our original target selection was performed before the release of Gaia DR2 astrometric data, it is reasonable to expect a modest to significant level of field contamination in our sample. This is confirmed in Fig. \ref{fig:obs_fields} which shows the Gaia EDR3 \citep[EDR3;][]{2016A&A...595A...1G,2020arXiv201201533G,2020arXiv201203380L} PMs of our targets overlaid on those of stars within the Jacobi radius of Pal 5. The astrometric data complements the radial velocities and [Fe/H] measurements that we will present in this paper and in Sec. \ref{kin} we will detail how we combine all this information to isolate a clean sample of Pal 5 stars. 

\subsection{Reduction \& Processing}
We began by reducing the spectra with 2df Data Reduction (2\textsc{dfdr}\footnote{\url{http://www.aao.gov.au/2df/aaomega/aaomega_2dfdr.html}}) software package, with the default settings for both gratings. This performed  the standard processes of debiasing, flat-fielding, wavelength calibration, extraction and sky subtraction using the designated sky fibres. At the end of this process, we removed any cosmic ray residuals by median-combining the spectra for each star. In the regions of the \cat\ triplet, the signal-to-noise of our targets ranges from $\sim2-30$ per pixel in Field 1 and $\sim5-50$ per pixel for Field 2.  Due to their superior signal-to-noise and spectral resolution, we use these red arm spectra for the bulk of our analysis.

To determine the radial velocities (RVs) of our targets, we used the python package \textsc{PyAstronomy}\footnote{https://github.com/sczesla/PyAstronomy} \citep{pya}. The \texttt{crosscorrRV} routine takes a target spectrum and cross-correlates it with a template spectrum, returning the shift as a velocity measurement. The template used was a synthetic spectrum that contained the \cat\ triplet lines broadened by a Gaussian at the resolution of the 1700D filter using the \texttt{instrBroadGaussFast} routine. We added random noise to each pixel in the target spectra based on the variance of the flux in that pixel \citep[e.g.,][]{2018MNRAS.477.4565S}, and subsequently performed the cross-correlation across the wavelength interval of 8450 \r{A}-8700 \r{A} - a region mostly bereft of sky residuals. This process was repeated 100 times per star to produce a distribution of measured radial velocities. We fit a Gaussian to the resulting distribution and present the mean and standard deviation as the measured radial velocity and associated 1$\sigma$ uncertainty. Lastly, we corrected for barycentric motion using the routine \texttt{helcorr}, and from now on we refer to these heliocentric radial velocities as $V_{R}$. 

We also use the \cat\ triplet lines to determine stellar metallicities through the well-established method based on their equivalent widths (EWs) \citep[e.g.,][]{1991AJ....101.1329A,2010A&A...513A..34S}. Using the \texttt{equivalent\_width} routine in the \textsc{specutils}\footnote{https://specutils.readthedocs.io/en/stable/} python package, we measured the EWs of all three of the \cat\ triplet lines on normalised spectra that had been wavelength-shifted to zero heliocentric velocity.  Specifically, we used this routine to fit a Gaussian profile in 10 \r{A} bandpasses centred on each \cat\ line.    In order to estimate the EW uncertainties for a given star, we repeated these measurements  on each of the 100 realizations created in the radial velocity calculations. Similarly, we adopt the mean and sigma of a Gaussian fit to the resultant distribution as the EW and its associated uncertainty. 

To derive the metallicity of an RGB star, the summed \cat\ EW,  $\Sigma EW_{\ion{Ca}{II}}$, can be related to [Fe/H] through knowledge of either the distance to the star or the difference between the V magnitude of the star and that of HB of the Pal 5 cluster, $V-V_{HB}$.  In this work, we adopt the  \cat\ calibration presented in \citet{2013MNRAS.434.1681C} which is valid over the range $-4.0\leq$ [Fe/H]$ \leq +0.5$.  This calibration takes the form of:

\begin{equation}
\begin{split}
{\rm[Fe/H]}=a+b\times (V-V_{HB})+c\times \Sigma EW_{\ion{Ca}{II}}\\+d\times \Sigma EW_{\ion{Ca}{II}}^{-1.5}+e\times \Sigma EW_{\ion{Ca}{II}} \times (V-V_{HB})
\label{eq:feh}
\end{split}
\end{equation}
where the coefficients $a,\,b,\,c,\,d$ and $e$ are listed in Table. \ref{tab:coefflist}. We adopt $V_{HB}$=17.51 mag for Pal 5 \citep{1996AJ....112.1487H}  and calculated the V-band magnitudes of our stars from their PS1 photometry using the transformation equations in \cite{2018BlgAJ..28....3K}.    Uncertainties in the metallicities come from combining the uncertainties of the \cat\ EWs and the uncertainties on the calibration coefficients.  Because of the assumption of a fixed magnitude for the HB, it should be emphasised that the metallicities we derive are strictly valid for genuine RGB stars at the distance of Pal 5.  While \citet{2016ApJ...819....1I} detect a slight distance gradient along the Pal~5 stream ranging from $\Delta(m-M)= 0.14\pm0.09$ mag at the extent of the trailing edge to $-0.09\pm0.04$ mag at the extent of the leading edge, this shift in distance translates to a mere $\sim0.03$~dex in [Fe/H] and is well within our uncertainties. 

Our blue arm spectra have sufficient resolution to individually measure the gravity-sensitive \ion{Mg}{I} triplet lines at $\sim$5170~\r{A}, which have been shown to be useful  for separating foreground dwarfs from the RGB stars that we are interested in \cite[e.g.,][ K15]{2009AJ....137.3378O}.  The EWs of these lines have been measured in the same way as the \cat\ triplet lines described above.  While the EW of the stronger \ion{Mg}{I} 8807~\r{A} line could also have been used for dwarf-giant separation \citep[e.g.,][]{Battaglia2012},  it was not always available due to flexure in the spectrograph. 

\begin{table}
\centering

\caption{List of coefficients from Table 4 from \citet{2013MNRAS.434.1681C} used in Eq. \ref{eq:feh}.}. 
\label{tab:coefflist}
\begin{tabular}{@{}cc@{}}
\hline \hline
Coefficient & Value\\
\hline
a&$-3.45\pm0.04$\\
b&$0.11\pm0.02$\\
c&$0.44\pm0.006$\\
d&$-0.65\pm0.12$\\
e&$0.03\pm0.003$\\
\hline
\end{tabular}
\end{table}

\subsection{Identifying Pal 5 stream stars} \label{subsec:mem} 

The distance and faintness of the Pal~5 stream conspire to make it challenging to cleanly isolate member stars from the significant foreground and background contaminant populations along its sightline.  This is further exacerbated by the presence of the Sgr stream in this part of the sky \citep{2016ApJ...819....1I, 2020ApJ...889...70B}. While the Sgr stream lies at a larger line-of-sight distance, red clump stars from this system contaminate the region of the CMD where faint Pal 5 RGB stars lie \citep[see Fig. 1 of][]{2020ApJ...889...70B}. These considerations motivate us to pursue a probabilistic approach to membership assignment in which we combine the new spectroscopic information described above with Gaia EDR3 astrometry and PS1 photometry.

We began by cross-matching our observed targets with EDR3, matching stars with the closest EDR3 source within a search radius of 2 arcsec.  We then removed any stars with a well resolved parallax, $(\omega - 3\sigma_{\omega})>0$ mas, since these will lie in the foreground. We also disregarded stars with a measured velocity uncertainty $>5$ km s$^{-1}$, as this value corresponds to spectra with poor signal-to-noise. Together, these cuts remove 434 stars (61 per cent) of the total sample. To avoid any potential biases, we do not consider further cuts on PMs or line-of-sight velocities at this stage of analysis.

Our approach to determining the probability of a given star belonging to the Pal 5 stream ($P_{P5}$), or to the field ($P_{MW}$), is to use the log-likelihood ($\ln{\mathcal{L}_{R}}$) ratio, otherwise known as the Neyman-Pearson test (see chapter 9 of \citealt{1993stp..book.....L}).  To this end, we consider in turn the likelihood ratio of the following features: dwarf/giant separation ($\mathcal{L}_{R,D/G}$), [Fe/H] ($\mathcal{L}_{R,[Fe/H]}$) and proximity to the dereddened Pal 5 RGB $(\mathcal{L}_{R,CMD})$. The combined log-likelihood is:

\begin{equation}\label{eq:loglike_tot}
\ln{\mathcal{L}_{R}}=\ln{\mathcal{L}_{R,CMD}} + \ln{\mathcal{L}_{R,D/G}}+\ln{\mathcal{L}_{R,[Fe/H]}}
\end{equation}

\noindent where $\mathcal{L}_{R,(CMD,D/G,[Fe/H])}$ is the ratio $P_{P5}/P_{MW}$ for the specific feature. A $\ln{\mathcal{L}_{R}}$>0 implies that a star has a higher likelihood of being a Pal 5 stream member,  while $\ln{\mathcal{L}_{R}}$<0 implies it is more likely to belong to the contaminating field. A value of $\ln{\mathcal{L}_{R}}$=0 implies neither scenario is favoured.

\subsubsection{Photometric Selection}\label{subsubsec:photsel}
Fig. \ref{fig:obs_fields} shows that our target selection broadly traces the RGB of Pal 5. To quantify the probability that a given star belongs to Pal 5 based on its CMD position, we measure its difference in colour from a Dartmouth stellar isochrone selected to best represent the cluster \citep{2008ApJS..178...89D}. For this, we use an isochrone 
with [Fe/H]=$-1.56$ dex \citep[as measured by ][]{2017A&A...601A..41K}) and age$=13.5$ Gyr, shifted to the distance of Pal 5.

Firstly, we de-reddened the PS1 photometry for each target using the updated \citet{1998ApJ...500..525S} reddening maps from \citet{2011ApJ...737..103S} with the Python package, DustMaps\footnote{\url{https://dustmaps.readthedocs.io/en/latest/index.html}} \citep{2018JOSS....3..695M}. The de-reddened magnitudes are denoted $g_0$ and $i_0$. We then assigned each star a probability according to the following equation \citep[e.g.,][]{2019MNRAS.485.2010G}:

\begin{equation}
P_{P5,CMD}=\frac{1}{\sqrt{2 \pi \sigma_{(g-i)_{0}}^2}} \exp \left(\frac{-((g-i)_{0}-(g-i)_{iso})^2}{2\sigma^2_{(g-i)_{0}}}\right)\label{eq:p5cmd}
\end{equation}

\noindent where $(g-i)_0$ and $(g-i)_{iso}$ are the de-reddened colour and isochrone colour at the star's $i_0$-band magnitude, and 
 $\sigma_{(g-i)_0}$ is the uncertainty in the de-reddened colour. 
 In this and the following steps, we have only considered stars that lie along the RGB; targets that are located along the HB will be dealt with separately (see Sec. \ref{kin}). 

The colour width of our RGB selection box (see Fig. \ref{fig:obs_fields})
is at most $\Delta(g-i)_0=0.4$ for a given $i_0$ mag. Across this rather modest range in colour, we have assumed for simplicity that field contaminants are uniformly distributed and assign each star a probability based on a uniform probability distribution. That is:

\begin{equation}
P_{MW,CMD} = 1/\Delta(g-i)_0\label{eq:fieldcmd}
\end{equation}

 Combining the probabilities from equations \ref{eq:p5cmd} and \ref{eq:fieldcmd}, we define $\mathcal{L}_{R,CMD}=P_{P5,CMD}/P_{MW,CMD}$.

\subsubsection{Dwarf/Giant Separation}

The gravity-sensitive \ion{Mg}{I} lines are commonly used to 
distinguish between RGB stars and foreground dwarf stars.  For a given metallicity and temperature, these features have larger EWs in high-gravity dwarf stars than they do in giants. 
The \ion{Mg}{I} lines around 5170 \r{A} have been used for this purpose in previous studies of the Pal 5 stream \cite[e.g.,][ K15]{2009AJ....137.3378O} as well as other diffuse halo structures \citep{2012AJ....143...88C}. These latter authors explicitly demonstrated that this is an effective discriminant for giant stars at the metallicity of Pal 5 and a population of more metal-rich dwarfs.

\begin{figure}
  \begin{center}  
    \includegraphics[width=\columnwidth]{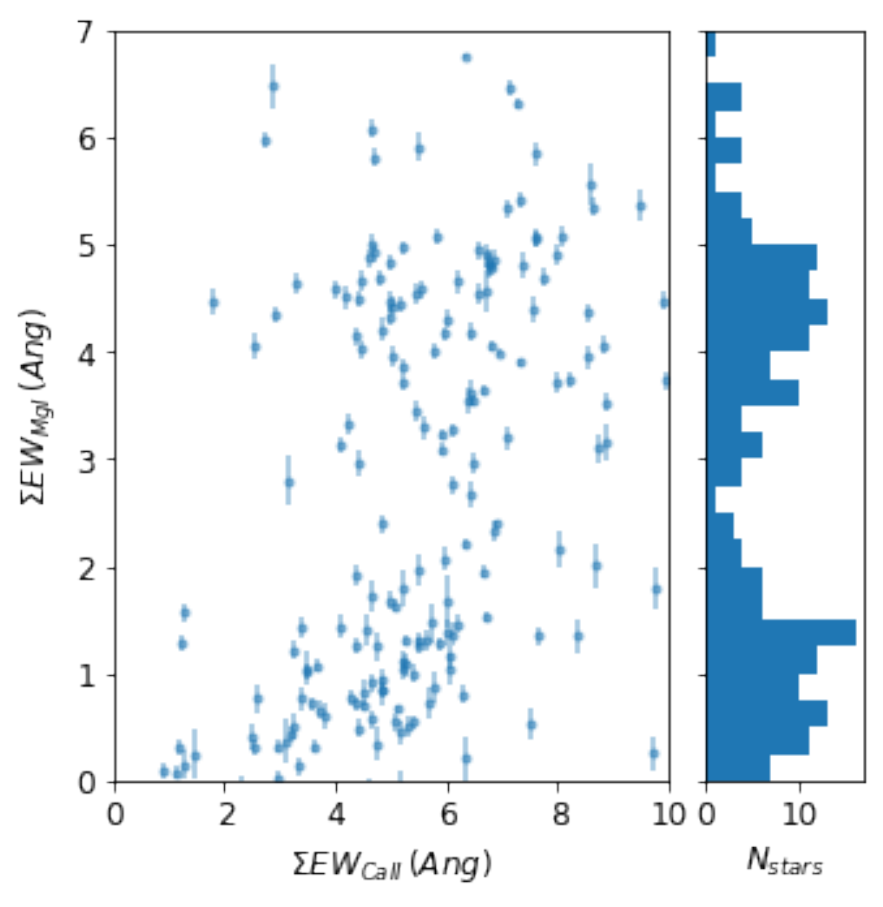}
  \end{center}
\caption{Dwarf/Giant separation for our targets after the removal of targets with significant EDR3 parallax measurements and low signal-to-noise spectra. The sum of the \cat\ EWs is shown along the x-axis, while the  summed \ion{Mg}{I} b EWs are along the y-axis. The right histogram shows the distribution of summed \ion{Mg}{I} b EWs when integrating over the x-axis. Two peaks can be clearly seen, corresponding to the dwarf population (peak near $\Sigma EW_{\mathrm{\ion{Mg}{I}}}$=5) and the giant population (lower peak near $\Sigma EW_{\mathrm{\ion{Mg}{I}}}=1$).}
\label{fig:DGS_dist}
\end{figure}

In Fig. \ref{fig:DGS_dist} we plot the summed EWs of the \ion{Mg}{I} and  \cat\ triplets measured for our sample. This diagram shows bimodal structure with the field dwarfs populating the higher sums of the \ion{Mg}{I} EW at a given \cat\ EW (see also K15). The one-dimensional distribution of \ion{Mg}{I} triplet EWs ($\Sigma EW_{\mathrm{{\ion{Mg}{I}}}}$) is shown the right panel and can be modelled by two normal distributions with the form:

\begin{equation}\label{eq:dgs}
P_{X,D/G} = \frac{1}{\sqrt{2 \pi(\sigma_{\Sigma EW_{\mathrm{\ion{Mg}{I}}}}^2+\sigma_{X}^2}) }
\exp \left(\frac{-(\Sigma EW_{\mathrm{\ion{Mg}{I}- X)^2}}}{2 (\sigma_{\Sigma EW_{\mathrm{\ion{Mg}{I}}}}^2+\sigma_{X}^2)}\right)
\end{equation}

\noindent where $X$ and $\sigma_X$ are the mean and standard deviations of the dwarf and giant distributions, and $\sigma_{\Sigma EW_{\mathrm{\ion{Mg}{I}}}}$ is the uncertainty in $\Sigma EW_{\mathrm{\ion{Mg}{I}}}$. After both distributions have been found, we calculate the likelihood ratio of $\mathcal{L}_{R,D/G}=P_{P5,D/G}/P_{MW,D/G}$.

\subsubsection{Metallicity Selection}
\begin{figure}
  \begin{center}  
    \includegraphics[width=\columnwidth]{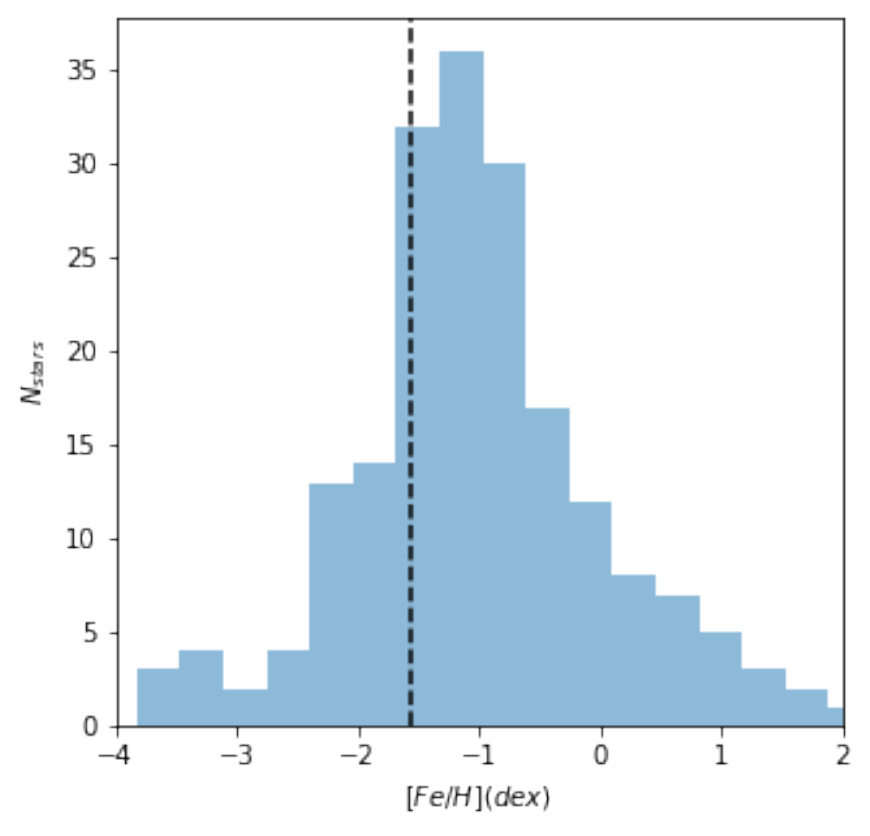}
  \end{center}
\caption{Distribution of \cat\ EW inferred [Fe/H] for all observed targets,  after the removal of targets with significant EDR3 parallax measurements and low signal-to-noise spectra. These values only reflect genuine [Fe/H] for stars at the distance of Pal 5. The dashed line indicates the measured [Fe/H] of Pal 5 (-1.56 dex) from high resolution spectroscopy.}
\label{fig:FEH}
\end{figure}

\citet{2017A&A...601A..41K} presented a detailed chemical analysis of Pal 5 based on high resolution spectra of 15 RGB stars. They derive a mean metallicity of [Fe/H]$_{P5}= -1.56\pm0.02\pm0.06$ dex (presented with statistical and systematic uncertainties). As there is no evidence for any [Fe/H] spread in the main body of Pal 5, we assume that the stars in the tidal stream will posses the same [Fe/H] abundance. Therefore, we have adopted this value when assigning stars their associated $P_{P5,[Fe/H]}$, which takes the form of another normal distribution:

\begin{equation}\label{eq:metp5}
\begin{aligned}
P_{P5,[Fe/H]}=\frac{1}{\sqrt{2 \pi (\sigma^2_{[Fe/H]}+\sigma^2_{[Fe/H]_{P5}})}}\cross \\
\exp \left(\frac{-([Fe/H]-[Fe/H]_{P5})^2}{2(\sigma^2_{[Fe/H]}+\sigma^2_{[Fe/H]_{P5}})}\right)
\end{aligned}
\end{equation}

\noindent where [Fe/H] is our measured value using the \cat\ EW, $\sigma_{[Fe/H]}$ is its uncertainty and $\sigma_{[Fe/H]_{P5}}$ is the uncertainty of $[Fe/H]_{P5}$, obtained by summing the statistical and systematic terms in  quadrature. 
To inform our choice of a model for the field component, we plot in Fig. \ref{fig:FEH} the inferred [Fe/H] distribution of all stars which remain after parallax and low signal-to-noise removal. While these values only represent actual metallicities for stars at the distance of Pal 5, it is notable that the overall distribution of this quantity can be well-described by a normal distribution. Therefore, we can define $P_{MW,[Fe/H]}$  in a similar manner to Eq. \ref{eq:metp5}:

\begin{equation}
\begin{aligned}
P_{MW,[Fe/H]}=\frac{1}{\sqrt{2 \pi ({\sigma^2_{[Fe/H]}+\sigma^2_{[Fe/H]_{MW}}})}}\cross \\
exp\left(\frac{-([Fe/H]-[Fe/H]_{MW})^2}{2(\sigma^2_{[Fe/H]}+\sigma^2_{[Fe/H]_{MW}})}\right)\label{eq:metMW}
\end{aligned}
\end{equation}
\noindent where $[Fe/H]_{MW}$ and $\sigma_[Fe/H]_{MW}$ are the mean and sigma of the normal distribution fit to the data. We note that while a small number of confirmed Pal 5 stream members will be included in our model of the field component, they are greatly outnumbered by the much more significant contaminant population, making any bias in determining $P_{MW,[Fe/H]}$ unlikely. Finally, we can calculate the likelihood ratio as $\mathcal{L}_{R,[Fe/H]}=P_{P5,[Fe/H]}/P_{MW,[Fe/H]}$.

\subsubsection{Calculating Membership Probability}
To transform our log-likelihoods into membership probabilities, we can calculate the probability that a given star belongs to Pal~5 ($P_{mem}$) using the following equation:

\begin{equation}\label{eq:loglike1}
P_{mem}=\frac{f\mathcal{L}_{P5}}{f \mathcal{L}_{P5} + (1-f) \mathcal{L}_{MW}}
\end{equation}

\noindent where $\mathcal{L}_{P5/MW}$ is the likelihood value with respect Pal~5 and the field respectively (not to be confused with the likelihood ratio $\mathcal{L}_{R}$), and $f$ is the normalisation value between the two populations. In our case, we have simply assumed that membership to Pal 5 or the MW field is equally probable, thus $f=0.5$. We acknowledge that this is arbitrary, however it does remove any bias towards either Pal 5 membership or the field which is reasonable in our analysis. As we have calculated the log-likelihood ratio ($\ln{\mathcal{L}_{R}}$) from Eq. \ref{eq:loglike_tot}, we can simplify Eq. \ref{eq:loglike1} by removing the common factor of $0.5\mathcal{L}_{MW}$ from the denominator to define the probability of a star that belongs to Pal 5 in terms of the likelihood ratio $\mathcal{L}_{R}$ defined in Eq. \ref{eq:loglike_tot}, \cite[e.g.][]{1993stp..book.....L,2011MNRAS.413.2895J}:

\begin{equation}\label{eq:loglike_P5}
P_{mem
}=\frac{\mathcal{L}_{R}}{\mathcal{L}_{R}+ 1}
\end{equation}

As we have set the normalisation factor to 0.5, this implies that stars with $P_{mem,P5} \geq 0.5$ are high likelihood Pal 5 members. With this assumption, we find 75 stars or $\sim11$\ per cent of our total sample are likely to belong to Pal 5. To avoid introducing biases, we have thus far ignored the kinematics of stars in identifying likely Pal 5 stream members. In the next Section, we will now consider how such measurements can inform our final sample definition. 

\begin{figure*}
  \begin{center}  
    \includegraphics[width=0.9\textwidth]{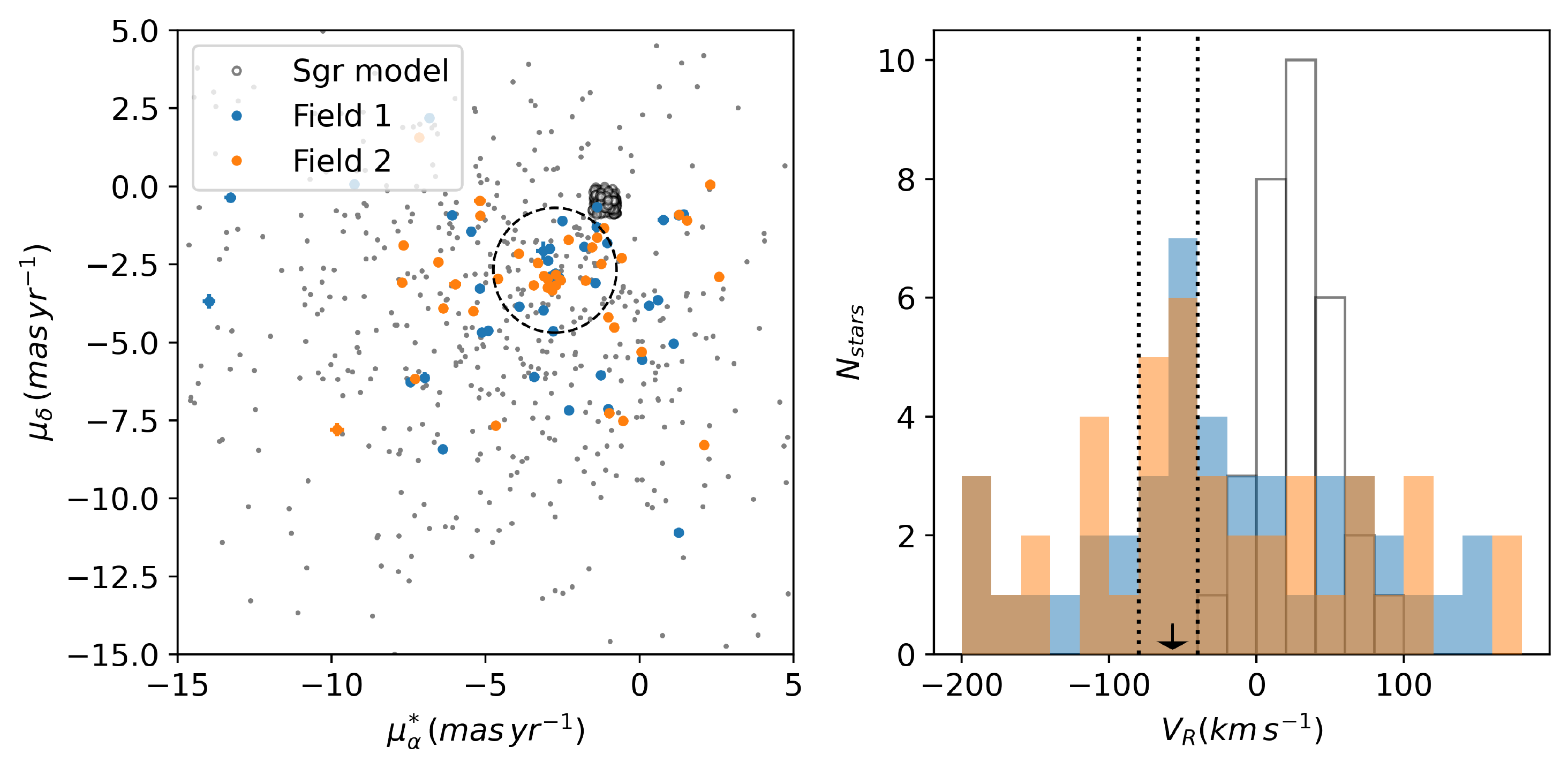}
  \end{center}
\caption{Diagnostic plots of the velocity measurements used in our membership assignment. Stars with $P_{mem}>=0.5$ are colour-coded according to field location, as in Fig. \ref{fig:obs_fields}, while stars with $P_{mem}<0.5$ are shown in grey.  Left: Proper motion distribution, with the dashed ring indicating the 2 mas yr$^{-1}$ boundary used for final sample definition. The open grey circles indicate the expected proper motion of Sgr stars from the model presented in \citet{2021MNRAS.501.2279V}, which lie outside the the dashed circle.  Right: Radial velocity histogram with the vertical dashed lines indicating our selected velocity range of --80 to --40 km s$^{-1}$. The solid arrow shows the radial velocity of Pal 5 at --57.4 km s$^{-1}$. Also shown by the grey unfilled histogram is the expected velocity distribution of Sgr stars from the model presented in \citet{2021MNRAS.501.2279V}, which are all outside our velocity range.}
\label{fig:kin_select}
\end{figure*}

\subsection{Incorporating Kinematics for Final Sample Definition}\label{kin}
 
The PM of Pal 5 has been measured repeatedly using Gaia DR2 and EDR3 data \citep[e.g.,][]{2019MNRAS.484.2832V,2021MNRAS.505.5978V}. 
However, the area of interest in this study is located outside the main body of the cluster and any PM filtering for our sample needs to account for how the PM of stars change as a function of position along the stream.   This has recently been examined by \cite{2019AJ....158..223P} who find that while the PMs of stars do vary along the length of the stream, they do not substantially change over the regions we are probing in this study.   As a result, we proceed with stars that lie within 2 mas yr$^{-1}$ of the main body PM of Pal 5\footnote{$\mu^{*}_{\alpha}=\mu_{\alpha}\cos{\delta}$}: $(\mu^{*}_{\alpha},\mu_{\delta})=(-2.75,-2.65)$ mas yr$^{-1}$ \citep{2019MNRAS.484.2832V}. This is a more conservative estimate than the 3 mas yr$^{-1}$ threshold used by  \citet{2020MNRAS.493.4978S}, which is large enough to include contamination from the Sgr stream. Indeed, using the model of \cite{2021MNRAS.501.2279V} in the region 223$\degr<\alpha<229.5\degr$ deg and $-20\degr<\delta< 10\degr$, we find that the Sgr stream in this direction has a PM of $(\mu^{*}_{\alpha},\mu_{\delta})\approx(-1.1,-0.5)$ mas yr$^{-1}$ (see Fig. \ref{fig:kin_select}). Our PM selection thus provides a cleaner sample of stars in the Pal 5 leading tail but may potentially suffer from minor  incompleteness if there are very energetic stars present. Disentangling such a hot population from the contamination from the Sgr stream and the Milky Way field will be very difficult based on currently-available data but may be possible with either improved distances and/or detailed chemistry (e.g. elemental abundance ratios).

The radial velocity distribution of our high likelihood Pal 5 members is also shown in Fig. \ref{fig:kin_select}. This shows an obvious grouping of stars between $-80$ and $-40$ km s$^{-1}$ but with many outliers. The \cite{2021MNRAS.501.2279V} Sgr stream model predicts that stream stars along this sightline will contribute at radial velocities $\geq -20$ km s$^{-1}$ and so do not pose an issue. As previous kinematic studies have found a radial velocity gradient of $\sim -1$ km s$^{-1}$ deg$^{-1}$ along the stream (e.g. K15, Ib17), we find it reasonable to expect that Pal 5 stream members will be contained within this velocity range across the radial extent of our fields. 

When our constraints on PMs and radial velocities are incorporated into our sample definition, this removes all but 15 of the high probability Pal 5 stars.  To this sample, we also add one star located on the HB (i.e. with $(g-i)_{0}\le 0.6$ mag in Fig. \ref{fig1}).  The HB was not considered as part of our photometric selection described in Sec. \ref{subsubsec:photsel} but we expect there to be very little field contamination in this part of the CMD. For this population, we have only considered the kinematic cuts in the PM and radial velocity, and find that these are satisfied by a single star in the outermost field. In total, we present these 16 stars as {\it bona fide} Pal 5 stream members and their properties are listed in Table \ref{tab:p5stars}. 
 
\begin{table*}
\centering

\caption{List of {\it bona fide} Pal 5 stream members from our AAT sample. The naming convention indicates with field the star belongs to, followed by its designated number.}
\label{tab:p5stars}
\begin{tabular}{@{}ccccccc@{}}
\hline \hline
Star&R.A.&Dec&$V_R$&$(g-i)_{0}$&$i_{0}$&[Fe/H]\\
(Field-number)&(deg, J2000)&(deg, J2000)& ($\rm{km}\,\rm{s}^{-1}$)&\multicolumn{2}{c}{(mag)}&(dex)\\
\hline
1-448$^{*}$&228.419&-0.708&$-56.87\pm1.98$&0.83&16.3&$0.01\pm0.13$\\
1-396&228.539&-0.878&$-65.50\pm3.91$&0.69&17.8&$-3.04\pm1.04$\\
1-272$^{*}$&228.118&-1.214&$-56.09\pm1.87$&0.68&17.7&$-0.59\pm0.41$\\
1-265&227.972&-1.245&$-51.52\pm2.13$&0.73&16.7&$-1.59\pm0.16$\\
1-190$^{*}$&227.663&-1.446&$-58.37\pm0.96$&0.85&16.2&$-0.99\pm0.11$\\
1-225$^{*}$&227.840&-1.484&$-52.00\pm1.44$&0.77&16.9&$-1.42\pm0.14$\\
1-135&227.327&-1.637&$-46.83\pm1.55$&0.81&16.7&$-0.82\pm0.12$\\
1-113&227.168&-1.760&$-40.43\pm4.36$&0.70&17.3&$-0.72\pm0.15$\\
2-346&224.939&-4.332&$-59.97\pm1.74$&0.69&17.0&$-1.22\pm0.1$\\
2-312&224.645&-4.525&$-62.05\pm1.63$&0.72&17.0&$-1.06\pm0.09$\\
2-253&223.607&-4.805&$-79.72\pm2.04$&0.78&16.9&$-1.28\pm0.11$\\
2-214&225.323&-4.819&$-67.20\pm1.17$&0.78&16.4&$-1.99\pm0.07$\\
2-550$^{\dagger}$&224.349&-5.052&$-49.54\pm2.92$&-0.35&17.4&$-1.96\pm0.1$\\
2-132&224.982&-5.087&$-48.56\pm3.7$&0.62&17.8&$-3.22\pm0.16$\\
2-069&224.783&-5.326&$-70.84\pm0.28$&0.89&15.7&$-1.47\pm0.04$\\
2-033&224.177&-5.501&$-63.32\pm1.15$&0.78&16.5&$-1.15\pm0.08$\\
\hline
 \multicolumn{7}{l}{$^{*}$ These stars are in common with Ib17 - the velocity presented here is the weighted-average value.}\\
 \multicolumn{7}{l}{$^{\dagger}$ Horizontal branch star.}\\
\end{tabular}

\end{table*}

\begin{figure*}
  \begin{center}  
    \includegraphics[width=0.8\textwidth]{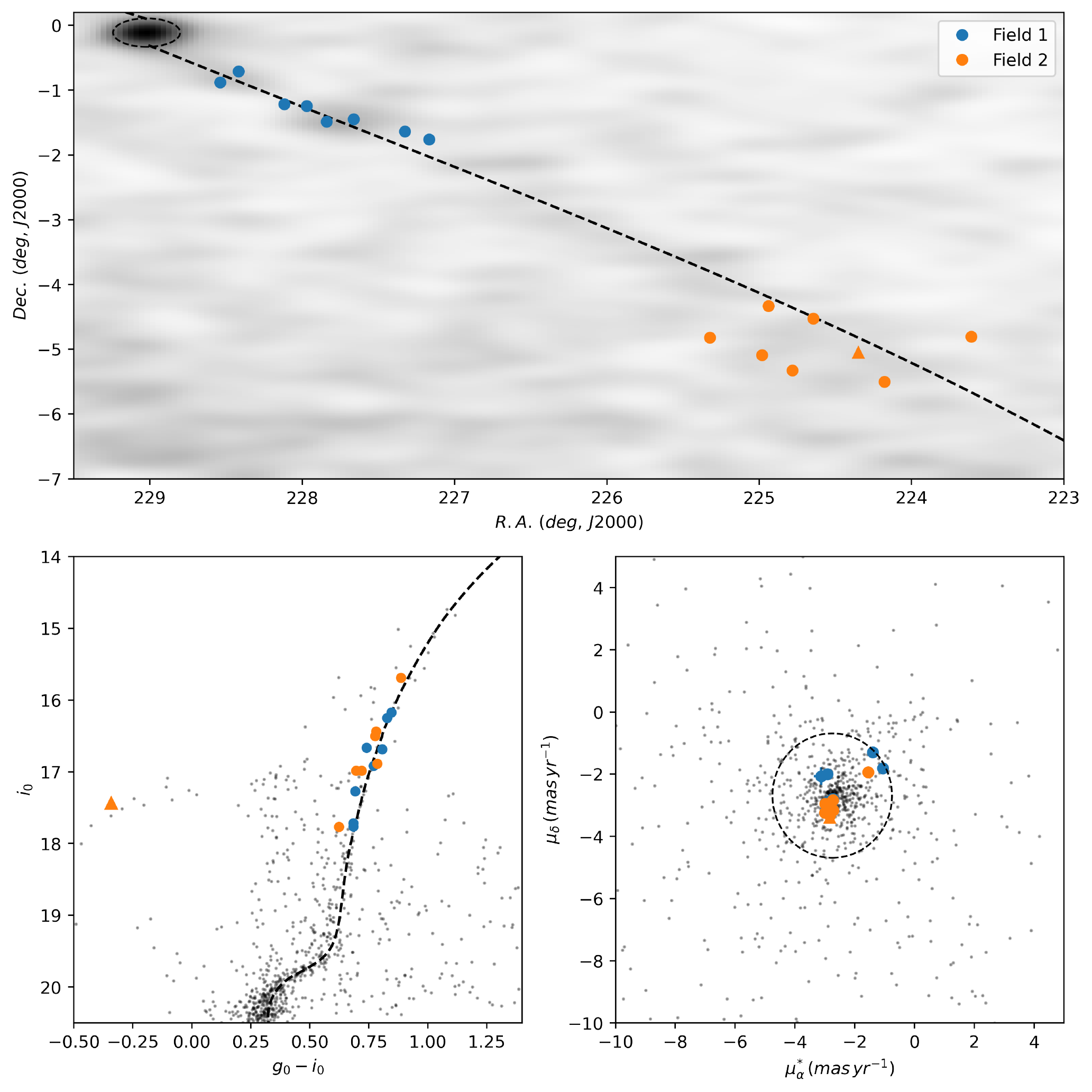}
  \end{center}
\caption{Our {\it bona fide} Pal 5 stars. Stars are colour-coded by which field they are located in, as in Fig. \ref{fig:obs_fields}. Top row: On-sky distribution of our member  stars. Stars in the inner field closely follow the stream track whereas those in the outermost field display a broader distribution on the sky. Bottom left: CMD of member stars. Also plotted is the Dartmouth stellar isochrone used in the photometric selection described in section \ref{subsubsec:photsel}. Bottom right: PM distribution of member stars.
}
\label{fig:evidence}
\end{figure*}

\section{Results}
In Fig. \ref{fig:evidence}, we show the spatial distribution, CMD and PM distribution of our final sample of Pal 5 stream stars. We see that eight of these stars lie in Field 1, the closest to the cluster centre, and they tightly follow the stream track. The remaining eight stars lie in the outermost field and appear to show a significant spread in position on the sky, suggestive of stream fanning; these stars lie further along the leading tail than has been previously explored spectroscopically. 

To be able to better quantify this behaviour, and also examine how our new measurements compare to the known gradients along the stream, we combine
our radial velocities with literature measurements to revisit the kinematics of the Pal 5 stream along its entire observed extent. Before doing this, we note that of the eight stars we have confirmed in Field 1, four of these stars have been previously identified by Ib17: stars 1-190, 1-225, 1-277 and 1-448. We compare the radial velocities that we have derived for these stars with those found by Ib17 and find an offset of $V_R-V_{Ib17}=0.4 \pm 1.9$ km~s$^{-1}$. As Ib17's measurements are based on higher signal-to-noise spectra than ours, we adopt a weighted-average between Ib17's and our measured velocity for these stars for the rest of the analysis (see Table \ref{tab:p5_fullist}).

\begin{figure*}
  \begin{center}  
    \includegraphics[width=0.9\textwidth]{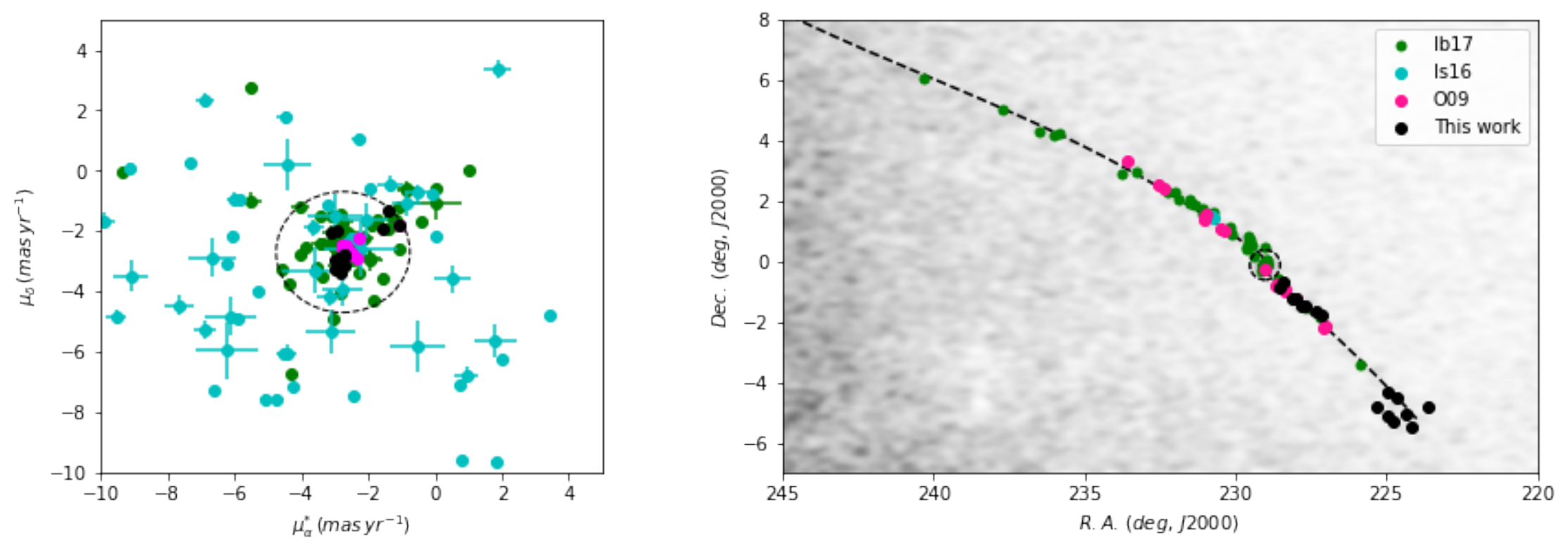}
  \end{center}
\caption{Left: EDR3 PM distribution of Ib17 (green), \citet{2016ApJ...823..157I} (light blue) and \citet{2009AJ....137.3378O} (pink) {\it bona fide} Pal 5 members. Our sample are shown as black points.  The black dashed ring displays a 2 mas yr$^{-1}$ radius around the measured Pal 5 PM from \citet{2021MNRAS.505.5978V}. It is clear that significant contamination exists in these earlier samples. Right: Spatial distribution of in our combined sample of \pk{109} {\it bona fide} stream members with the stream track of \pk{\citet{2016ApJ...819....1I} and \citet{2017ApJ...842..120I} under-laid}. The stars closely follow the stream track until the edge of the leading tail, near ($\alpha, \delta$)$\sim (225,-5)$, where the fanning reported in this paper is observed. The dashed circle indicates the location of Pal~5 itself.}
\label{fig:all_spat_pm}
\end{figure*}

With the astrometric measurements provided by EDR3 now available, we are also able to  clean previous kinematic samples of Pal~5 stars from contaminants. We have gathered the radial velocities that have been presented in \cite{2009AJ....137.3378O} (18 stars), \cite{2016ApJ...823..157I} (130 stars) and Ib17 (130 stars), noting that the latter includes all stars that were previously reported in K15 and 13 stars from \cite{2009AJ....137.3378O}. We subject these samples to the same astrometric and kinematic selections we performed on our own sample - that is,  removing stars with a resolved parallax $(\omega - 3\sigma_{\omega})>0$ mas,  radial velocity uncertainties $\geq$ 5 km s$^{-1}$, radial velocities outside the range of $-80$ and $-40$ km s$^{-1}$ and PMs larger than 2 mas yr$^{-1}$ of the main body of the Pal 5 cluster. 

As seen in the left panel in Fig. \ref{fig:all_spat_pm}, there is significant contamination in these earlier kinematic samples of Pal 5 stream stars.   We find that out of the 130 stars from Ib17  \citep[including the stars stars in common with our sample and][]{2009AJ....137.3378O}, only 91 stars (75 unique, 16 in common with other data sets) stars satisfy the astrometric  and kinematic selection criteria adopted in this paper. For the \cite{2016ApJ...823..157I} sample, only 2 out of the 130 stars are retained and  for the \cite{2009AJ....137.3378O} sample, 
16 out of 18 stars are retained (of which 12 are present in Ib17). Accounting for stars in common and unique to each sample, we present four, two and 75 unique stars (81 unique stars in total) as Pal 5 system members from the studies of  \cite{2009AJ....137.3378O}, \cite{2016ApJ...823..157I} and Ib17 respectively. Adding in the 12 stars in common between \cite{2009AJ....137.3378O} and Ib17, and our 16 stars (of which four are in common with Ib17), we present an extended sample of 109 {\it bona fide} members belonging to the Pal 5 cluster and its tidal tails stars spread across $\sim$22$\degr$ on the sky. The properties of these stars, including their sky positions, radial velocities, PMs and PS1 photometry, are presented in Table \ref{tab:p5_fullist} while their spatial distribution across the sky is shown in the right panel of Fig \ref{fig:all_spat_pm}.

\begin{figure*}
  \begin{center}  
    \includegraphics[width=0.8\textwidth]{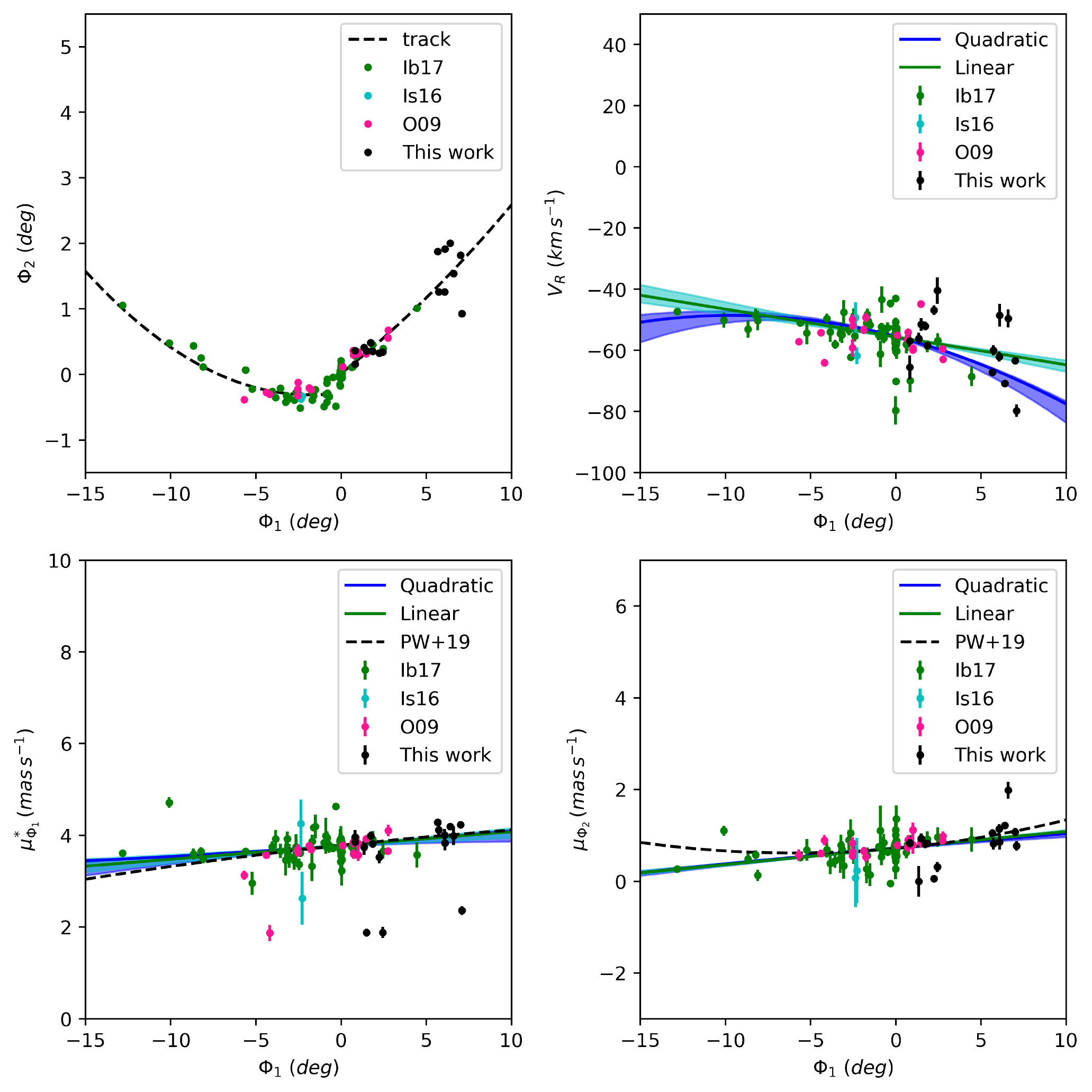}
  \end{center}
\caption{Inferred trends in various parameters along the stream, using the fit parameters from Table \ref{tab:kin_models} displayed as a function of the stream coordinate $\phi_{1}$. Top left: sky positions in the Pal 5 stream coordinate frame $(\phi_{1},\phi_{2})$. Top right: Radial velocities along the stream. Bottom left: $\mu_{\phi_{1}}$ along the stream. Bottom right: $\mu_{\phi_{2}}$ along the stream. The linear and quadratic model fits to the radial velocities and proper motions are showin in green and blue respectively, with the shaded regions indicating the 1$\sigma$ uncertainties on the corresponding model. The points are the same as Fig. \ref{fig:all_spat_pm} (left), and the dashed black line indicates the quadratic fit to the RRL PMs from \citet{2019AJ....158..223P}.}
\label{fig:kim_fig}
\end{figure*}

The next stage in our analysis is to transform the PMs and spatial coordinates of this extended sample to a spherical coordinate system aligned with the stream. For this, we adopt the transformation provided by \cite{2020ApJ...889...70B} in which the cluster origin lies at $(\phi_{1}, \phi_{2})=(0,0)\degr$ and the leading arm of the tidal tail is along the positive direction.  We fit a second-order polynomial to these data to calculate the best fit track to the trailing and leading tails. We find that the trailing tail can be described as:

\begin{equation}
\phi_{2,trailing}(\phi_1)=0.0108\phi^2_{1}+0.0394\phi_{1}-0.2768
\end{equation}

\noindent while the leading tail follows: 

\begin{equation}
\phi_{2,leading}(\phi_1)=0.0053\phi^2_{1}+0.2043\phi_{1}+0.0163
\end{equation}

\noindent in stream coordinates (see top left panel of Fig. \ref{fig:kim_fig}). In constructing these fits, we have excluded all stars that lie within the Jacobi radius to avoid the cluster influencing the track. 

\begin{figure}
  \begin{center}  
    \includegraphics[width=0.8\columnwidth]{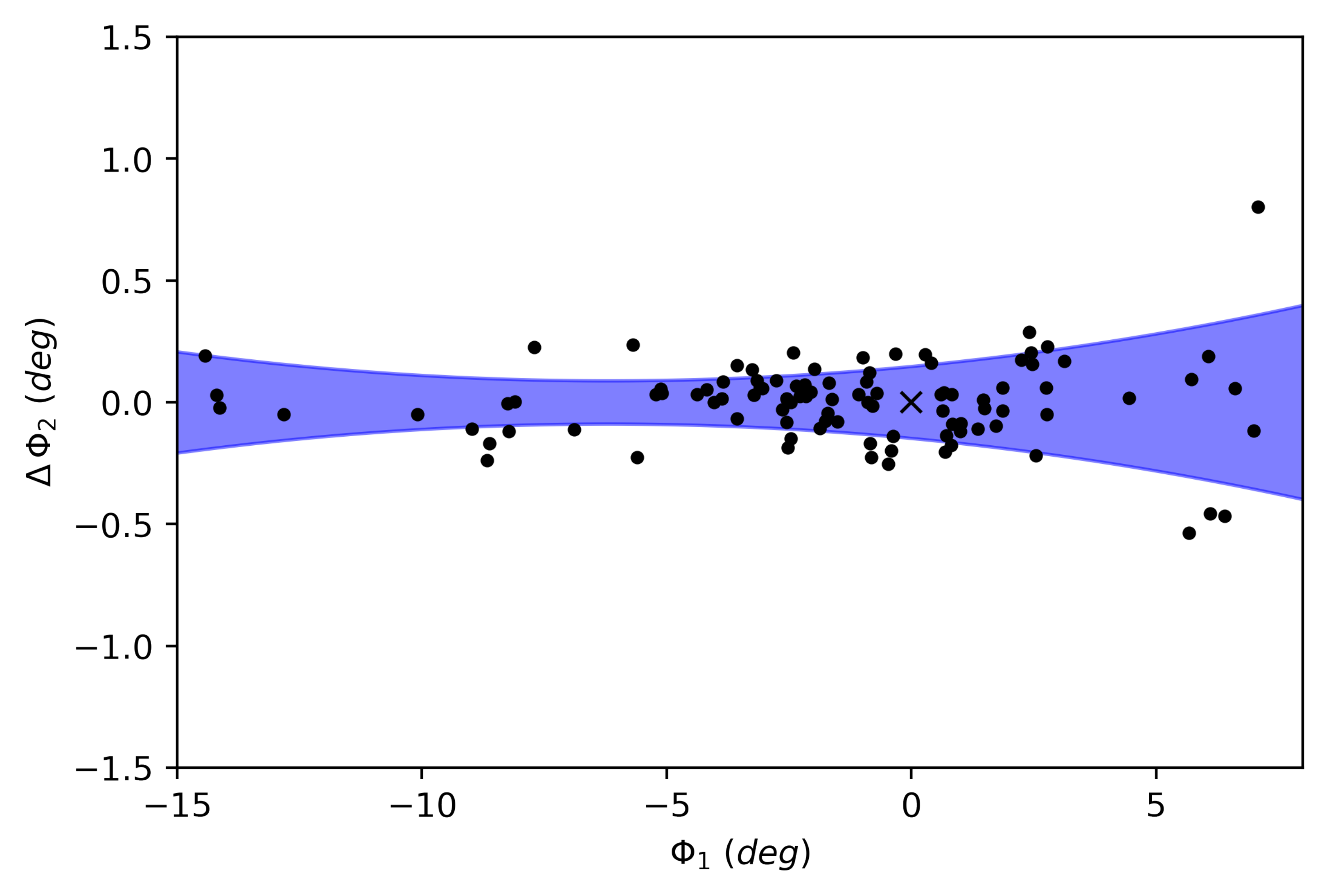}
  \end{center}
\caption{Stream width (blue shaded region) as a function of position along the Pal 5  stream. The black points represent stars in the extended sample. The large increase in width in the leading tail, beyond $\phi_1 > 5 \degr$,  is indicative of the fanning. The black cross indicates the position of Pal 5 itself.}
\label{fig:fanning}
\end{figure}

To characterise the stream width along its length, we calculate for each star the difference between its coordinate $\phi_2$ and the stream track at its $\phi_1$ position, which we denote as $\Delta\phi_2$. We then fit the mean $\Delta\phi_2$ (in $2.5\degr$ bins as a function of $\phi_1$) with a second order polynomial to estimate the behaviour of the width along the stream (Fig. \ref{fig:fanning}).  We see an increase in stream width at $\phi_{1}>5\degr$ reaching approximately 0.5$\degr$ at  $\phi_{1}=7\degr$, confirming the visual impression of fanning in the leading tail that is seen in Fig. \ref{fig:evidence}. This also agrees well with a photometric detection of fanning in this region presented by \cite{2020ApJ...889...70B} (see their Fig. 3). The stream width becomes smaller, as expected, as we move along the stream towards the trailing tail and up until $\phi_{1}= -10 \degr$ can be characterised by a width of $\approx 0.2\degr$.  The polynomial fit suggests that the stream may become wider again at the observed edge of the trailing tail but there are too few stars in this region for this to be meaningful. 

\begin{table*}
\centering

\caption{List of Pal 5 members compiled in this study. These stars are drawn from the new data presented in this work (K22) as well those from the previous studies of  \citealt{2009AJ....137.3378O} (O09), \citet{2016ApJ...823..157I} (Is16) and Ib17 that are retained after astrometric and kinematic cleaning. We provide sky positions, PMs and their uncertainties, radial velocities and their uncertainties and the  $g$-band magnitudes and $(g-i)$ colours from PS1. The last column indicates the origin of their first identification. Only the top five rows are shown here, the rest is available online.}
\label{tab:p5_fullist}

\begin{tabular}{@{}ccccccccccc@{}}
\hline \hline
  \multicolumn{1}{c}{R.A.} &
  \multicolumn{1}{c}{Dec.} &
  \multicolumn{1}{c}{$\mu^{*}_{\alpha}$} &
  \multicolumn{1}{c}{$\sigma_{\mu^{*}_{\alpha}}$} &
  \multicolumn{1}{c}{$\mu_{\delta}$} &
  \multicolumn{1}{c}{$\sigma_{\mu_{\delta}}$} &
  \multicolumn{1}{c}{$V_R$} &
  \multicolumn{1}{c}{$\sigma_{V_R}$} &
  \multicolumn{1}{c}{$g_{PS1}$} &
  \multicolumn{1}{c}{$(g-i)_{PS1}$} &
  \multicolumn{1}{c}{Source} \\
\multicolumn{2}{c}{(deg, J2000)} &
\multicolumn{2}{c}{(mas yr$^{-1}$)} &
\multicolumn{2}{c}{(mas yr$^{-1}$)} &
\multicolumn{2}{c}{(km s$^{-1}$)} &
\multicolumn{2}{c}{(mag)}&\\
\hline
240.342&6.036&-2.89&0.03&-2.18&0.02&-47.35&0.78&16.18&1.16&Ib17\\
237.723&5.027&-3.35&0.12&-3.50&0.10&-50.13&2.50&18.47&0.83&Ib17\\
236.502&4.285&-2.74&0.13&-2.38&0.11&-53.06&2.87&18.66&0.93&Ib17\\
235.844&4.241&-2.84&0.11&-2.02&0.11&-50.22&3.22&18.04&0.87&Ib17\\
236.025&4.193&-2.74&0.10&-2.49&0.09&-48.50&2.16&18.16&0.84&Ib17\\

\hline
\end{tabular}

\end{table*}

With our extended Pal 5 sample, we can also revisit the radial velocity gradient across $\approx22\degr$ on the sky, compared to the 16$\degr$-18$\degr$ of the  \citet{2016ApJ...823..157I} and Ib17 spectroscopic samples.   Following Ib17, we have employed a ``conservative formulation" of a Bayesian automatic outlier rejection algorithm to fit the data \citep[see Chapter 8.3 of][]{Sivia:2006ty}. Using this technique, we explored both a linear ($A+ B \phi_{1}$) and quadratic ($A+B \phi_{1} + C\phi_{1}^2$) fit to the radial velocities and each component of the PMs as function of position along the stream,  $\phi_1$. The parameters of these fits are shown in Table \ref{tab:kin_models} and the models, along with their associated $1\sigma$ uncertainties, are shown in Fig. \ref{fig:kim_fig}.

For the linear model, we find a radial velocity gradient of $-0.81 \pm 0.14$ km s$^{-1}$ deg$^{-1}$ along the stream, which is
consistent within the uncertainties of the values found in other work where linear models have been adopted \citep[e.g.,][K15]{2009AJ....137.3378O,2016ApJ...823..157I}. When considering the radial velocity gradient of Ib16, 0.7 km s$^{-1}$ deg$^{-1}$, this gradient was calculated with respect to the standard gnomonic coordinate $\xi$ - a tangential horizontal projection about the cluster center. When we calculate our radial velocity gradient in that coordinate system, we find a  consistent value of $0.93 \pm 0.2$ km s$^{-1}$ deg$^{-1}$. No previous analyses have explored a quadratic fit to the stream radial velocity distribution along the tails. We find that the quadratic term is consistent with zero within uncertainties ($-0.07\pm 0.04$ km s$^{-1}$ deg$^{-1}$), while the linear term is similar ($-1.34\pm0.28$ km s$^{-1}$ deg$^{-1}$) to the aforementioned linear gradient within the uncertainties. We conclude that a linear model of the radial velocities 
remains the best description over the angular extent of the stream probed here. 

At $\phi_{1}=0$ (i.e. the location of the cluster), we find $V_{R}=-55.67\pm0.28$ km s$^{-1}$ with the linear model and $V_{R}=-55.50 \pm 0.28$ km s$^{-1}$ with the quadratic model, both similar to Ib17, though slightly different from the radial velocities of $-57.4\pm0.4$ km s$^{-1}$ from K15 and $-58.6\pm0.2$ km s$^{-1}$ from \citet{2019MNRAS.482.5138B}. We note that restricting the constant term in the linear and quadratic models to this velocity does not affect the results significantly. 

We perform similar fits to the PM components,  $\mu^*_{\phi_{1}}$\footnote{$\mu^*_{\phi_{1}}=\mu_{\phi_{1}}\cos{\phi_{2}}$} and $\mu_{\phi_{2}}$. In both instances, there is a small linear gradient of $0.03\pm0.01$ mas yr$^{-1}$ deg $^{-1}$ and $0.04\pm0.004$ mas yr$^{-1}$ deg$^{-1}$ for $\mu^{*}_{\phi_{1}}$ and $\mu_{\phi_{2}}$ respectively. Further, both leading terms in the quadratic fit to  $\mu^{*}_{\phi_{1}}$ and $\mu_{\phi_{2}}$, are consistent with zero within 1 $\sigma$.  \cite{2019AJ....158..223P} have recently studied the PMs of {\it bona fide} RRL candidates in Pal 5 as function of $\phi_{1}$ using a quadratic model. For the proper motion of the Pal 5 cluster, we find $(\mu^{*}_{\phi_{1}},\mu_{\phi_{2}})=(3.77\pm0.02,0.73\pm0.02)$ mas yr $^{-1}$, which is in excellent agreement with their value, as well as that of \citet{2021MNRAS.505.5978V} when transformed back to equatorial coordinates.  The PM trends we find along the length of the stream are in good agreement as well with their results (see the lower panels of Fig. \ref{fig:kim_fig}), which is very encouraging given that these trends have been fit to completely independent samples of stars. However, we note that
 while our linear and quadratic fits to $\mu_{\phi_{2}}$ are in very good agreement to theirs at locations $-10^{\circ} \la \phi_{1} \la 5^{\circ}$, they rapidly diverge from their quadratic model at the extreme ends  of the stream. The reason for this 
discrepancy is likely due to the relative lack of stars in these outermost reaches in the two samples -- \cite{2019AJ....158..223P} fit their model to 27 RRLs along the cluster/stream, with only a small number of stars guiding the fit at distances $|\phi_{1}|>10^{\circ}$. We, too, suffer from small number of stars at those locations along the stream. Clearly, it is essential to identify further {\it bona fide} members at such high distance from the progenitor to provide more stringent constraints on the 3D kinematics, as well as to establish if the fanning uncovered here continues further along the leading tail. 

Finally, we can also re-examine the radial velocity dispersion within the tails using our combined dataset. To do this, we define $\Delta V_R$ to be the difference between the observed radial velocity of a given star and the expected radial velocity at the star's location from the model fits in Table \ref{tab:kin_models}. We then fit the resulting distribution of $\Delta V_R$ with normal distribution with a mean of zero, and report the 1$\sigma$ value as the velocity dispersion, $\sigma$. Both the linear and quadratic model fits are found to show a similar radial velocity dispersion within the debris, $\sigma_{V}=2.17 \pm 0.27$ km s$^{-1}$ and $2.50 \pm 0.24$ km s$^{-1}$ respectively, which is consistent with previously published values (e.g. K15). We explored if there is any trend in the radial velocity dispersion along the tails and found that it that varies very little, increasing by $\lesssim 1$ km s$^{-1}$ into the fanned region of the leading tail. Further, we also find a low dispersion for both directions of proper motion: $\sigma_{\mu_{\phi_{1}}} = 0.12 \pm 0.01$ mas yr$^{-1}$, and $\sigma_{\mu_{\phi_{2}}} = 0.16 \pm 0.01$ mas yr$^{-1}$ within the tail, also with no significant variation along it. Hence, the velocity dispersion, along all three velocity dimensions, is characteristically low and rather constant within the current sample of stars. This is not surprising, as the Pal 5 stream has long been noted to be a kinematically cold structure \citep[e.g.,][]{2009AJ....137.3378O}.

\begin{table*}

\begin{center}
\caption{The coefficients of the linear ($A+ B \phi_{1}$) and quadratic ($A+B \phi_{1} + C\phi_{1}^2$) fits to the radial velocity $V_{R}$, and to the proper motions $\mu_{\Phi_{1}}$ and $\mu_{\Phi_{2}}$, as a function of the tangential projection along the stream, $\phi_{1}$. The top rows contain the fit coefficients while $\sigma$ shows the velocity and PM dispersion about the stream. }

\label{tab:kin_models}
\begin{tabular}{@{}ccccccc}

\hline \hline
Parameters&\multicolumn{2}{c}{$V_R$}&\multicolumn{2}{c}{$\mu^{*}_{\Phi_{1}}$}&\multicolumn{2}{c}{$\mu_{\Phi_{2}}$}\\
Model&linear  & quadratic&linear  & quadratic&linear  & quadratic\\
\hline
$A$&$-55.67\pm0.27$&$-55.50\pm0.28$&$3.77\pm0.02$&$3.77\pm0.02$&$0.73\pm0.02$&$0.73\pm0.02$\\
$B$&$-0.81\pm0.14$&$-1.34\pm0.28$&$0.03\pm0.01$&$0.03\pm0.01$&$0.04\pm0.004$&$0.03\pm0.004$\\
$C$&$--$&$-0.07\pm0.04$&$--$&$-0.0004\pm0.0005$&$--$&$-0.0003\pm0.0003$\\
$\sigma$&$2.17\pm0.27$&$2.75\pm0.24$&$0.12\pm0.01$&$0.13\pm0.01$&$0.16\pm0.01$&$0.16\pm0.01$\\
\hline

\end{tabular}

\end{center}
\end{table*}

\section{Discussion}

Prior to this work, spectroscopic studies of Pal 5 member stars have been confined to the most dense and identifiable regions of the tidal stream. However, in recent years, photometric studies have shown that there is a faint extension to the leading edge of the stream \citep{2016MNRAS.463.1759B}, and that the stars in this region may actually fan out in angular distance from the nominal stream track \citep{2020ApJ...889...70B}. The outermost field observed in the present study covers this previously-unexplored region and confirms the existence of stars in this region with kinematics and metallicities that make them highly probable to be {\it bona fide} members of the Pal 5 stream.  

Stream fanning has been explored from a theoretical standpoint by various authors.  \citet{2015ApJ...799...28P} showed how the morphology of the Pal 5 stream on its own could constrain the shape of the Milky Way's dark matter halo.  In particular, they argued against the triaxial potential of the \citet{2010ApJ...714..229L} for the dark halo on the basis of the thin, curved appearance of the extent of the Pal 5 stream known at that time.  Our confirmation of fanning in a more distant part of the leading tail is intriguing in this context but needs to be assessed against the fact that no such behaviour is seen along the well-studied trailing tail, as expected in their models.  \citet{2016ApJ...824..104P} discuss fanning in the Ophiuchus stream as a result of chaotic orbits caused by a rotating Galactic bar.  
The modelling work of \citet{2020ApJ...889...70B} shows that a  rotating massive bar can also lead to fanning in the leading tail of Pal 5 but that this scenario predicts other effects which are not compatible with the observations. 

Much of the recent modelling work on the Pal 5 stream has focused on the effect of other perturbations acting on the tails, seeking explanations for the observed inhomogeneous density structure, 
the marked length asymmetry of the leading and trailing arms and ``wiggles" in the stream track. 
Specifically, dark matter subhalos piercing the stream have been invoked to explain peaks and gaps in the density distribution \citep[e.g.][]{2012ApJ...760...75C}, and dynamical resonances induced by the presence of the Galactic bar have been exploited to explain the apparent truncation in the leading tail  
\citep[e.g.,][]{2017MNRAS.470...60E,2017NatAs...1..633P}. As shown by  \citet{2020ApJ...889...70B}, there is currently no one scenario that can explain all of the observations to date. The new radial velocities we provide here, as well as our construction of a PM-cleaned catalogue of previously-identified stream members, should prove very useful in future modelling work. 
	
Broadly speaking, we find that the radial velocities of the stars identified in the present study are in qualitative agreement with the predictions of the modelling work by \citet{2017MNRAS.470...60E}. However, it can be seen that in their work, which considers the evolution of the tails within a smooth static potential as well as ones in which interactions with dark matter clumps, giant molecular clouds and a rotating bar take place, the different models are largely degenerate in their leading arm radial velocity predictions (see their Figs. 7, 9 and 11). 

In this context, we point out that the internal structural and dynamical properties of the progenitor cluster may have equally important effects on the morphology and kinematics of the resulting tidal streams.  Indeed, \citet{2012MNRAS.420.2700K} showed that fan-like structures naturally form in stellar streams produced by a star cluster on an eccentric orbit within a tidal field.  These features result entirely from the expected dynamical evolution of a collisional stellar system, without the need to invoke any additional external perturbation. Further to this, it has been noted that unusual stream features, such as the "dog-leg" feature seen in the stream that wraps NGC 1097 \citep{2015arXiv150403697A}, can be reproduced only when a rotating progenitor is considered. Similarly, some of the features of the Sgr stream appear to confirm the presence of significant angular momentum in Sgr itself \citep{2021ApJ...908..244D}, and the N-body models that connect the massive GC NGC 5139 ($\omega$ Centauri) and the associated tidal stream Fimbulthul \citep{2019NatAs.tmp..258I}, require a progenitor with some level of internal rotation. These examples underscore the need for renewed efforts to explore how Pal 5 internal structure could affect the morphology, asymmetry and kinematics of its tidal tails, alongside further work on the role of external perturbations.

\section{Conclusions}
In this paper, we present the results of a spectroscopic study of the leading tail of the Pal 5 stream. We combine our measurements of line-strengths and radial velocities with Gaia EDR3 astrometry and PS1 photometry to derive membership probabilities following a 
log-likelihood ratio approach.  We find 16 stars with {\it bona fide} of membership, of which four were previously known and eight of them lie in a previously unexplored part of the leading tail.  The sky locations of these stars confirm the presence of fanning in this part of the stream, as recently suggested by \citet{2020ApJ...889...70B}.

We also revisit previous radial velocity studies of the Pal 5 stream and clean them of contaminants using astrometry measurements from Gaia ERD3. Combined with the new measurements presented in this paper, this yields a sample of  109 {\it bona fide} members of the Pal 5 GC and its tidal tails. We fit the radial velocities and PMs of this sample as a function of the stream coordinate $\phi_1$ with linear and quadratic models.  We find a linear radial velocity gradient of $-0.81 \pm 0.14$ km s$^{-1}$ deg$^{-1}$ across the extent of the stream, in keeping with previous studies that probed a more limited angular extent.  We also find a stream velocity dispersion of $2.17\pm0.27$ km s$^{-1}$, also in agreement with previous results. The quadratic fits across the debris return very similar results. 

We provide our catalogue in the hopes that it can be used for future modeling work and stress the importance of such work examining the role of both internal and external factors on the stream properties.  Most work to date has focused on external factors alone (e.g. dark matter sub-halo impacts, the influence of a rotating bar) but thus far none of these scenarios can explain the entirety of the current set of observations. 

Ultimately,  more observations of this iconic disrupting star cluster system are required. Deep photometry beyond the currently-known stream extent will allow for constraints to be placed on the presence of fanning in the trailing tail, and on whether the leading tail really does peter out in a low surface brightness fan or if it reappears again at more southern declinations \citep[e.g.][]{2017NatAs...1..633P}.  Further spectroscopic  measurements, especially in large areas adjacent to and beyond the currently-known extent of the tails, will also be critical for confirming membership in these extremely contaminated parts of the sky and for mapping out radial velocity and velocity dispersion trends. It is fortuitous that Pal 5's equatorial location means that it will be an accessible target for both the upcoming WEAVE \citep{2012SPIE.8446E..0PD} and 4MOST \citep{2019Msngr.175....3D} large multi-object spectrograph surveys, which are set to begin operations within the next 2-3 years.

\section*{Acknowledgements}
This work makes use of the following software packages: \textsc{astropy} \citep{2013A&A...558A..33A,2018AJ....156..123A}, \textsc{dustmaps} \citep{2018JOSS....3..695M}, \textsc{Gala} \citep{gala,adrian_price_whelan_2020_4159870}, \textsc{matplotlib} \citep{Hunter:2007}, \textsc{numpy} \citep{2011CSE....13b..22V}, \textsc{PyAstronomy} \citep{pya}, \textsc{scipy} \citep{2020SciPy-NMeth}, \textsc{SpecUtils} \citep{2020zndo...3718589E}. 

PBK is grateful for the support by a Commonwealth Rutherford Fellowship from the Commonwealth Scholarship Commission in the UK. ALV and PBK acknowledge support from a UKRI Future Leaders Fellowship (MR/S018859/1). 

Based on data acquired at the Anglo-Australian Telescope under program A/2017A/102 (Opticon 17A/063). We acknowledge the traditional custodians of the land on which the AAT stands, the Gamilaraay people, and pay our respects to elders past and present.

This project has received funding from the European Union's Horizon 2020 research and innovation programme under grant agreement No 730890. This material reflects only the authors views and the Commission is not liable for any use that may be made of the information contained therein.

This work has made use of data from the European Space Agency (ESA) mission {\it Gaia} (\url{https://www.cosmos.esa.int/gaia}), processed by the {\it Gaia} Data Processing and Analysis Consortium (DPAC, \url{https://www.cosmos.esa.int/web/gaia/dpac/consortium}). Funding for the DPAC has been provided by national institutions, in particular the institutions participating in the {\it Gaia} Multilateral Agreement.\\
The Pan-STARRS1 Surveys (PS1) and the PS1 public science archive have been made possible through contributions by the Institute for Astronomy, the University of Hawaii, the Pan-STARRS Project Office, the Max-Planck Society and its participating institutes, the Max Planck Institute for Astronomy, Heidelberg and the Max Planck Institute for Extraterrestrial Physics, Garching, The Johns Hopkins University, Durham University, the University of Edinburgh, the Queen's University Belfast, the Harvard-Smithsonian Center for Astrophysics, the Las Cumbres Observatory Global Telescope Network Incorporated, the National Central University of Taiwan, the Space Telescope Science Institute, the National Aeronautics and Space Administration under Grant No. NNX08AR22G issued through the Planetary Science Division of the NASA Science Mission Directorate, the National Science Foundation Grant No. AST-1238877, the University of Maryland, Eotvos Lorand University (ELTE), the Los Alamos National Laboratory, and the Gordon and Betty Moore Foundation.
\textit{Facility}: AAT (AAOmega)

\section*{Data Availability}
The list of members of Pal~5 which underpins this article and corresponding figures are available in an extended version of Table 4, provided as online supplementary material.




\bibliographystyle{mnras}
\bibliography{bibfile} 

\begin{thebibliography}{}
\makeatletter
\relax
\def\mn@urlcharsother{\let\do\@makeother \do\$\do\&\do\#\do\^\do\_\do\%\do\~}
\def\mn@doi{\begingroup\mn@urlcharsother \@ifnextchar [ {\mn@doi@}
  {\mn@doi@[]}}
\def\mn@doi@[#1]#2{\def\@tempa{#1}\ifx\@tempa\@empty \href
  {http://dx.doi.org/#2} {doi:#2}\else \href {http://dx.doi.org/#2} {#1}\fi
  \endgroup}
\def\mn@eprint#1#2{\mn@eprint@#1:#2::\@nil}
\def\mn@eprint@arXiv#1{\href {http://arxiv.org/abs/#1} {{\tt arXiv:#1}}}
\def\mn@eprint@dblp#1{\href {http://dblp.uni-trier.de/rec/bibtex/#1.xml}
  {dblp:#1}}
\def\mn@eprint@#1:#2:#3:#4\@nil{\def\@tempa {#1}\def\@tempb {#2}\def\@tempc
  {#3}\ifx \@tempc \@empty \let \@tempc \@tempb \let \@tempb \@tempa \fi \ifx
  \@tempb \@empty \def\@tempb {arXiv}\fi \@ifundefined
  {mn@eprint@\@tempb}{\@tempb:\@tempc}{\expandafter \expandafter \csname
  mn@eprint@\@tempb\endcsname \expandafter{\@tempc}}}

\bibitem[\protect\citeauthoryear{{Ahn} et~al.,}{{Ahn}
  et~al.}{2012}]{2012ApJS..203...21A}
{Ahn} C.~P.,  et~al., 2012, \mn@doi [\apjs] {10.1088/0067-0049/203/2/21}, \href
  {https://ui.adsabs.harvard.edu/abs/2012ApJS..203...21A} {203, 21}

\bibitem[\protect\citeauthoryear{{Amorisco}, {Martinez-Delgado}  \&
  {Schedler}}{{Amorisco} et~al.}{2015}]{2015arXiv150403697A}
{Amorisco} N.~C.,  {Martinez-Delgado} D.,   {Schedler} J.,  2015, arXiv
  e-prints, \href {https://ui.adsabs.harvard.edu/abs/2015arXiv150403697A} {p.
  arXiv:1504.03697}

\bibitem[\protect\citeauthoryear{Armandroff \& Da~Costa}{Armandroff \&
  Da~Costa}{1991}]{1991AJ....101.1329A}
Armandroff T.~E.,  Da~Costa G.~S.,  1991, AJ, 101, 1329

\bibitem[\protect\citeauthoryear{Battaglia \& Starkenburg}{Battaglia \&
  Starkenburg}{2012}]{Battaglia2012}
Battaglia G.,  Starkenburg E.,  2012, \aap, 539, 123

\bibitem[\protect\citeauthoryear{Baumgardt, Hilker, Sollima  \&
  Bellini}{Baumgardt et~al.}{2019}]{2019MNRAS.482.5138B}
Baumgardt H.,  Hilker M.,  Sollima A.,   Bellini A.,  2019, MNRAS, 482, 5138

\bibitem[\protect\citeauthoryear{{Bernard} et~al.,}{{Bernard}
  et~al.}{2016}]{2016MNRAS.463.1759B}
{Bernard} E.~J.,  et~al., 2016, \mn@doi [\mnras] {10.1093/mnras/stw2134}, \href
  {https://ui.adsabs.harvard.edu/abs/2016MNRAS.463.1759B} {463, 1759}

\bibitem[\protect\citeauthoryear{Bonaca, Geha, K{\"u}pper, Diemand, Johnston
  \& Hogg}{Bonaca et~al.}{2014}]{2014ApJ...795...94B}
Bonaca A.,  Geha M.,  K{\"u}pper A. H.~W.,  Diemand J.,  Johnston K.~V.,   Hogg
  D.~W.,  2014, ApJ, 795, 94

\bibitem[\protect\citeauthoryear{Bonaca et~al.,}{Bonaca
  et~al.}{2020}]{2020ApJ...889...70B}
Bonaca A.,  et~al., 2020, ApJ, 889, 70

\bibitem[\protect\citeauthoryear{Bovy, Bahmanyar, Fritz  \& Kallivayalil}{Bovy
  et~al.}{2016}]{2016ApJ...833...31B}
Bovy J.,  Bahmanyar A.,  Fritz T.~K.,   Kallivayalil N.,  2016, ApJ, 833, 31

\bibitem[\protect\citeauthoryear{Carlberg, Grillmair  \& Hetherington}{Carlberg
  et~al.}{2012}]{2012ApJ...760...75C}
Carlberg R.~G.,  Grillmair C.~J.,   Hetherington N.,  2012, ApJ, 760, 75

\bibitem[\protect\citeauthoryear{Carrera, Pancino, Gallart  \& del
  Pino}{Carrera et~al.}{2013}]{2013MNRAS.434.1681C}
Carrera R.,  Pancino E.,  Gallart C.,   del Pino A.,  2013, MNRAS, 434, 1681

\bibitem[\protect\citeauthoryear{Casey, Keller  \& Da~Costa}{Casey
  et~al.}{2012}]{2012AJ....143...88C}
Casey A.~R.,  Keller S.~C.,   Da~Costa G.,  2012, AJ, 143, 88

\bibitem[\protect\citeauthoryear{Chambers et~al.,}{Chambers
  et~al.}{2016}]{2016arXiv161205560C}
Chambers K.~C.,  et~al., 2016, arXiv, p. arXiv:1612.05560

\bibitem[\protect\citeauthoryear{{Czesla}, {Schr{\"o}ter}, {Schneider},
  {Huber}, {Pfeifer}, {Andreasen}  \& {Zechmeister}}{{Czesla}
  et~al.}{2019}]{pya}
{Czesla} S.,  {Schr{\"o}ter} S.,  {Schneider} C.~P.,  {Huber} K.~F.,  {Pfeifer}
  F.,  {Andreasen} D.~T.,   {Zechmeister} M.,  2019, {PyA: Python
  astronomy-related packages} (\mn@eprint {ascl} {1906.010})

\bibitem[\protect\citeauthoryear{Dalton et~al.,}{Dalton
  et~al.}{2012}]{2012SPIE.8446E..0PD}
Dalton G.,  et~al., 2012, in McLean I.~S.,  Ramsay S.~K.,   Takami H.,  eds,
  Proceedings of the SPIE. SPIE, p. 84460P

\bibitem[\protect\citeauthoryear{Dehnen, Odenkirchen, Grebel  \& Rix}{Dehnen
  et~al.}{2004}]{2004AJ....127.2753D}
Dehnen W.,  Odenkirchen M.,  Grebel E.~K.,   Rix H.-W.,  2004, AJ, 127, 2753

\bibitem[\protect\citeauthoryear{{Dey} et~al.,}{{Dey}
  et~al.}{2019}]{2019AJ....157..168D}
{Dey} A.,  et~al., 2019, \mn@doi [\aj] {10.3847/1538-3881/ab089d}, \href
  {https://ui.adsabs.harvard.edu/abs/2019AJ....157..168D} {157, 168}

\bibitem[\protect\citeauthoryear{Dotter, Chaboyer, Jevremovi{\'c}, Kostov,
  Baron  \& Ferguson}{Dotter et~al.}{2008}]{2008ApJS..178...89D}
Dotter A.,  Chaboyer B.,  Jevremovi{\'c} D.,  Kostov V.,  Baron E.,   Ferguson
  J.~W.,  2008, ApJS, 178, 89

\bibitem[\protect\citeauthoryear{{Earl} et~al.,}{{Earl}
  et~al.}{2020}]{2020zndo...3718589E}
{Earl} N.,  et~al., 2020, {astropy/specutils: v1.0},
  \mn@doi{10.5281/zenodo.3718589}

\bibitem[\protect\citeauthoryear{Erkal, Koposov  \& Belokurov}{Erkal
  et~al.}{2017}]{2017MNRAS.470...60E}
Erkal D.,  Koposov S.~E.,   Belokurov V.,  2017, MNRAS, 470, 60

\bibitem[\protect\citeauthoryear{{Gaia Collaboration} et~al.,}{{Gaia
  Collaboration} et~al.}{2016}]{2016A&A...595A...1G}
{Gaia Collaboration} et~al., 2016, A{\&}A, 595, A1

\bibitem[\protect\citeauthoryear{{Gaia Collaboration} et~al.,}{{Gaia
  Collaboration} et~al.}{2018}]{2018A&A...616A...1G}
{Gaia Collaboration} et~al., 2018, A{\&}A, 616, A1

\bibitem[\protect\citeauthoryear{{Gaia Collaboration}, Brown, Vallenari,
  Prusti, de Bruijne, Babusiaux  \& Biermann}{{Gaia Collaboration}
  et~al.}{2020}]{2020arXiv201201533G}
{Gaia Collaboration} Brown A. G.~A.,  Vallenari A.,  Prusti T.,  de Bruijne J.
  H.~J.,  Babusiaux C.,   Biermann M.,  2020, arXiv, p. arXiv:2012.01533

\bibitem[\protect\citeauthoryear{Green}{Green}{2018}]{2018JOSS....3..695M}
Green G.,  2018, Journal of Open Source Software, 3, 695

\bibitem[\protect\citeauthoryear{Gregory, Collins, Read, Irwin, Ibata, Martin,
  McConnachie  \& Weisz}{Gregory et~al.}{2019}]{2019MNRAS.485.2010G}
Gregory A.~L.,  Collins M. L.~M.,  Read J.~I.,  Irwin M.~J.,  Ibata R.~A.,
  Martin N.~F.,  McConnachie A.~W.,   Weisz D.~R.,  2019, MNRAS, 485, 2010

\bibitem[\protect\citeauthoryear{Grillmair \& Dionatos}{Grillmair \&
  Dionatos}{2006}]{2006ApJ...641L..37G}
Grillmair C.~J.,  Dionatos O.,  2006, ApJ, 641, L37

\bibitem[\protect\citeauthoryear{Harris}{Harris}{1996}]{1996AJ....112.1487H}
Harris W.~E.,  1996, AJ, 112, 1487

\bibitem[\protect\citeauthoryear{Hunter}{Hunter}{2007}]{Hunter:2007}
Hunter J.~D.,  2007, Computing in Science and Engineering, 9, 90

\bibitem[\protect\citeauthoryear{Ibata, Lewis  \& Martin}{Ibata
  et~al.}{2016}]{2016ApJ...819....1I}
Ibata R.~A.,  Lewis G.~F.,   Martin N.~F.,  2016, ApJ, 819, 1

\bibitem[\protect\citeauthoryear{Ibata, Lewis, Thomas, Martin  \&
  Chapman}{Ibata et~al.}{2017}]{2017ApJ...842..120I}
Ibata R.~A.,  Lewis G.~F.,  Thomas G.,  Martin N.~F.,   Chapman S.,  2017, ApJ,
  842, 120

\bibitem[\protect\citeauthoryear{Ibata, Bellazzini, Malhan, Martin  \&
  Bianchini}{Ibata et~al.}{2019}]{2019NatAs.tmp..258I}
Ibata R.~A.,  Bellazzini M.,  Malhan K.,  Martin N.,   Bianchini P.,  2019,
  Nature Astronomy, 112, 1487

\bibitem[\protect\citeauthoryear{Ishigaki, Hwang, Chiba  \& Aoki}{Ishigaki
  et~al.}{2016}]{2016ApJ...823..157I}
Ishigaki M.~N.,  Hwang N.,  Chiba M.,   Aoki W.,  2016, ApJ, 823, 157

\bibitem[\protect\citeauthoryear{{Jenkins} \& {Peacock}}{{Jenkins} \&
  {Peacock}}{2011}]{2011MNRAS.413.2895J}
{Jenkins} C.~R.,  {Peacock} J.~A.,  2011, \mn@doi [\mnras]
  {10.1111/j.1365-2966.2011.18361.x}, \href
  {https://ui.adsabs.harvard.edu/abs/2011MNRAS.413.2895J} {413, 2895}

\bibitem[\protect\citeauthoryear{{\VAN{Jong}{De}{de}}~Jong
  et~al.,}{{\VAN{Jong}{De}{de}}~Jong et~al.}{2019}]{2019Msngr.175....3D}
{\VAN{Jong}{De}{de}}~Jong R.~S.,  et~al., 2019, The Messenger, 175, 3

\bibitem[\protect\citeauthoryear{Koch \& C{\^o}t{\'e}}{Koch \&
  C{\^o}t{\'e}}{2017}]{2017A&A...601A..41K}
Koch A.,  C{\^o}t{\'e} P.,  2017, A{\&}A, 601, A41

\bibitem[\protect\citeauthoryear{{Kostov} \& {Bonev}}{{Kostov} \&
  {Bonev}}{2018}]{2018BlgAJ..28....3K}
{Kostov} A.,  {Bonev} T.,  2018, Bulgarian Astronomical Journal, \href
  {https://ui.adsabs.harvard.edu/abs/2018BlgAJ..28....3K} {28, 3}

\bibitem[\protect\citeauthoryear{{K{\"u}pper}, {Lane}  \&
  {Heggie}}{{K{\"u}pper} et~al.}{2012}]{2012MNRAS.420.2700K}
{K{\"u}pper} A. H.~W.,  {Lane} R.~R.,   {Heggie} D.~C.,  2012, \mn@doi [\mnras]
  {10.1111/j.1365-2966.2011.20242.x}, \href
  {https://ui.adsabs.harvard.edu/abs/2012MNRAS.420.2700K} {420, 2700}

\bibitem[\protect\citeauthoryear{K{\"u}pper, Balbinot, Bonaca, Johnston, Hogg,
  Kroupa  \& Santiago}{K{\"u}pper et~al.}{2015}]{2015ApJ...803...80K}
K{\"u}pper A. H.~W.,  Balbinot E.,  Bonaca A.,  Johnston K.~V.,  Hogg D.~W.,
  Kroupa P.,   Santiago B.~X.,  2015, ApJ, 803, 80

\bibitem[\protect\citeauthoryear{Kuzma, Da~Costa, Keller  \& Maunder}{Kuzma
  et~al.}{2015}]{2015MNRAS.446.3297K}
Kuzma P.~B.,  Da~Costa G.~S.,  Keller S.~C.,   Maunder E.,  2015, MNRAS, 446,
  3297

\bibitem[\protect\citeauthoryear{{Kuzma}, {Ferguson}  \&
  {Pe{\~n}arrubia}}{{Kuzma} et~al.}{2021}]{2021MNRAS.507.1127K}
{Kuzma} P.~B.,  {Ferguson} A.~M.~N.,   {Pe{\~n}arrubia} J.,  2021, \mn@doi
  [\mnras] {10.1093/mnras/stab2280}, \href
  {https://ui.adsabs.harvard.edu/abs/2021MNRAS.507.1127K} {507, 1127}

\bibitem[\protect\citeauthoryear{Law \& Majewski}{Law \&
  Majewski}{2010}]{2010ApJ...714..229L}
Law D.~R.,  Majewski S.~R.,  2010, ApJ, 714, 229

\bibitem[\protect\citeauthoryear{{Lewis} et~al.,}{{Lewis}
  et~al.}{2002}]{2002MNRAS.333..279L}
{Lewis} I.~J.,  et~al., 2002, \mn@doi [\mnras]
  {10.1046/j.1365-8711.2002.05333.x}, \href
  {https://ui.adsabs.harvard.edu/abs/2002MNRAS.333..279L} {333, 279}

\bibitem[\protect\citeauthoryear{Lindegren et~al.,}{Lindegren
  et~al.}{2020}]{2020arXiv201203380L}
Lindegren L.,  et~al., 2020, arXiv, p. arXiv:2012.03380

\bibitem[\protect\citeauthoryear{{Lupton}}{{Lupton}}{1993}]{1993stp..book.....L}
{Lupton} R.,  1993, {Statistics in theory and practice}

\bibitem[\protect\citeauthoryear{Mastrobuono-Battisti, Di~Matteo, Montuori  \&
  Haywood}{Mastrobuono-Battisti et~al.}{2012}]{2012A&A...546L...7M}
Mastrobuono-Battisti A.,  Di~Matteo P.,  Montuori M.,   Haywood M.,  2012,
  A{\&}A, 546, L7

\bibitem[\protect\citeauthoryear{Niederste-Ostholt, Belokurov, Evans, Koposov,
  Gieles  \& Irwin}{Niederste-Ostholt et~al.}{2010}]{2010MNRAS.408L..66N}
Niederste-Ostholt M.,  Belokurov V.,  Evans N.~W.,  Koposov S.,  Gieles M.,
  Irwin M.~J.,  2010, MNRAS: Letters, 408, L66

\bibitem[\protect\citeauthoryear{Odenkirchen et~al.,}{Odenkirchen
  et~al.}{2001}]{2001ApJ...548L.165O}
Odenkirchen M.,  et~al., 2001, ApJ, 548, L165

\bibitem[\protect\citeauthoryear{Odenkirchen et~al.,}{Odenkirchen
  et~al.}{2003}]{2003AJ....126.2385O}
Odenkirchen M.,  et~al., 2003, AJ, 126, 2385

\bibitem[\protect\citeauthoryear{Odenkirchen, Grebel, Kayser, Rix  \&
  Dehnen}{Odenkirchen et~al.}{2009}]{2009AJ....137.3378O}
Odenkirchen M.,  Grebel E.~K.,  Kayser A.,  Rix H.-W.,   Dehnen W.,  2009, AJ,
  137, 3378

\bibitem[\protect\citeauthoryear{Pearson, K{\"u}pper, Johnston  \&
  Price-Whelan}{Pearson et~al.}{2015}]{2015ApJ...799...28P}
Pearson S.,  K{\"u}pper A. H.~W.,  Johnston K.~V.,   Price-Whelan A.~M.,  2015,
  ApJ, 799, 28

\bibitem[\protect\citeauthoryear{Pearson, Price-Whelan  \& Johnston}{Pearson
  et~al.}{2017}]{2017NatAs...1..633P}
Pearson S.,  Price-Whelan A.~M.,   Johnston K.~V.,  2017, Nature Astronomy, 1,
  633

\bibitem[\protect\citeauthoryear{Price-Whelan}{Price-Whelan}{2017}]{gala}
Price-Whelan A.~M.,  2017, \mn@doi [The Journal of Open Source Software]
  {10.21105/joss.00388}, 2

\bibitem[\protect\citeauthoryear{{Price-Whelan}, {Sesar}, {Johnston}  \&
  {Rix}}{{Price-Whelan} et~al.}{2016}]{2016ApJ...824..104P}
{Price-Whelan} A.~M.,  {Sesar} B.,  {Johnston} K.~V.,   {Rix} H.-W.,  2016,
  \mn@doi [\apj] {10.3847/0004-637X/824/2/104}, \href
  {https://ui.adsabs.harvard.edu/abs/2016ApJ...824..104P} {824, 104}

\bibitem[\protect\citeauthoryear{Price-Whelan, Mateu, Iorio, Pearson, Bonaca
  \& Belokurov}{Price-Whelan et~al.}{2019}]{2019AJ....158..223P}
Price-Whelan A.~M.,  Mateu C.,  Iorio G.,  Pearson S.,  Bonaca A.,   Belokurov
  V.,  2019, AJ, 158, 223

\bibitem[\protect\citeauthoryear{Price-Whelan et~al.,}{Price-Whelan
  et~al.}{2020}]{adrian_price_whelan_2020_4159870}
Price-Whelan A.,  et~al., 2020, adrn/gala: v1.3,
  \mn@doi{10.5281/zenodo.4159870}, \url
  {https://doi.org/10.5281/zenodo.4159870}

\bibitem[\protect\citeauthoryear{Schlafly \& Finkbeiner}{Schlafly \&
  Finkbeiner}{2011}]{2011ApJ...737..103S}
Schlafly E.~F.,  Finkbeiner D.~P.,  2011, ApJ, 737, 103

\bibitem[\protect\citeauthoryear{Schlegel, Finkbeiner  \& Davis}{Schlegel
  et~al.}{1998}]{1998ApJ...500..525S}
Schlegel D.~J.,  Finkbeiner D.~P.,   Davis M.,  1998, ApJ, 500, 525

\bibitem[\protect\citeauthoryear{{Sharp} et~al.,}{{Sharp}
  et~al.}{2006}]{2006SPIE.6269E..0GS}
{Sharp} R.,  et~al., 2006, in {McLean} I.~S.,  {Iye} M.,  eds,  Society of
  Photo-Optical Instrumentation Engineers (SPIE) Conference Series Vol. 6269,
  Society of Photo-Optical Instrumentation Engineers (SPIE) Conference Series.
  p. 62690G (\mn@eprint {arXiv} {astro-ph/0606137}), \mn@doi{10.1117/12.671022}

\bibitem[\protect\citeauthoryear{{Simpson}}{{Simpson}}{2018}]{2018MNRAS.477.4565S}
{Simpson} J.~D.,  2018, \mn@doi [\mnras] {10.1093/mnras/sty847}, \href
  {https://ui.adsabs.harvard.edu/abs/2018MNRAS.477.4565S} {477, 4565}

\bibitem[\protect\citeauthoryear{Sivia \& Skilling}{Sivia \&
  Skilling}{2006}]{Sivia:2006ty}
Sivia D.,  Skilling J.,  2006, {Data Analysis}.
A Bayesian Tutorial, Oxford University Press

\bibitem[\protect\citeauthoryear{Sollima}{Sollima}{2020}]{2020MNRAS.495.2222S}
Sollima A.,  2020, MNRAS, 495, 2222

\bibitem[\protect\citeauthoryear{Starkenburg et~al.,}{Starkenburg
  et~al.}{2010}]{2010A&A...513A..34S}
Starkenburg E.,  et~al., 2010, A{\&}A, 513, A34

\bibitem[\protect\citeauthoryear{Starkman, Bovy  \& Webb}{Starkman
  et~al.}{2020}]{2020MNRAS.493.4978S}
Starkman N.,  Bovy J.,   Webb J.~J.,  2020, MNRAS, 493, 4978

\bibitem[\protect\citeauthoryear{{The Astropy Collaboration} et~al.,}{{The
  Astropy Collaboration} et~al.}{2013}]{2013A&A...558A..33A}
{The Astropy Collaboration} et~al., 2013, A{\&}A, 558, A33

\bibitem[\protect\citeauthoryear{{The Astropy Collaboration} et~al.,}{{The
  Astropy Collaboration} et~al.}{2018}]{2018AJ....156..123A}
{The Astropy Collaboration} et~al., 2018, AJ, 156, 123

\bibitem[\protect\citeauthoryear{{Thomas}, {Ibata}, {Famaey}, {Martin}  \&
  {Lewis}}{{Thomas} et~al.}{2016}]{2016MNRAS.460.2711T}
{Thomas} G.~F.,  {Ibata} R.,  {Famaey} B.,  {Martin} N.~F.,   {Lewis} G.~F.,
  2016, \mn@doi [\mnras] {10.1093/mnras/stw1189}, \href
  {https://ui.adsabs.harvard.edu/abs/2016MNRAS.460.2711T} {460, 2711}

\bibitem[\protect\citeauthoryear{Vasiliev}{Vasiliev}{2019}]{2019MNRAS.484.2832V}
Vasiliev E.,  2019, MNRAS, 484, 2832

\bibitem[\protect\citeauthoryear{{Vasiliev} \& {Baumgardt}}{{Vasiliev} \&
  {Baumgardt}}{2021}]{2021MNRAS.505.5978V}
{Vasiliev} E.,  {Baumgardt} H.,  2021, \mn@doi [\mnras]
  {10.1093/mnras/stab1475}, \href
  {https://ui.adsabs.harvard.edu/abs/2021MNRAS.505.5978V} {505, 5978}

\bibitem[\protect\citeauthoryear{{Vasiliev}, {Belokurov}  \&
  {Erkal}}{{Vasiliev} et~al.}{2021}]{2021MNRAS.501.2279V}
{Vasiliev} E.,  {Belokurov} V.,   {Erkal} D.,  2021, \mn@doi [\mnras]
  {10.1093/mnras/staa3673}, \href
  {https://ui.adsabs.harvard.edu/abs/2021MNRAS.501.2279V} {501, 2279}

\bibitem[\protect\citeauthoryear{Virtanen et~al.,}{Virtanen
  et~al.}{2020}]{2020SciPy-NMeth}
Virtanen P.,  et~al., 2020, Nature Methods, 17, 261

\bibitem[\protect\citeauthoryear{{\VAN{Walt}{van}{van der}}~Walt, Colbert  \&
  Varoquaux}{{\VAN{Walt}{van}{van der}}~Walt
  et~al.}{2011}]{2011CSE....13b..22V}
{\VAN{Walt}{van}{van der}}~Walt S.,  Colbert S.~C.,   Varoquaux G.,  2011,
  arXiv, 13, 22

\bibitem[\protect\citeauthoryear{{del Pino}, {Fardal}, {van der Marel},
  {{\L}okas}, {Mateu}  \& {Sohn}}{{del Pino}
  et~al.}{2021}]{2021ApJ...908..244D}
{del Pino} A.,  {Fardal} M.~A.,  {van der Marel} R.~P.,  {{\L}okas} E.~L.,
  {Mateu} C.,   {Sohn} S.~T.,  2021, \mn@doi [\apj] {10.3847/1538-4357/abd5bf},
  \href {https://ui.adsabs.harvard.edu/abs/2021ApJ...908..244D} {908, 244}

\makeatother
\end{thebibliography}




\bsp	
\label{lastpage}
\end{document}